\documentclass{elsarticle}

\usepackage{soul}
\usepackage{subfigure}

\usepackage{amsmath}
\usepackage{amssymb}
\usepackage{hyperref}
\usepackage{color}
\usepackage{multirow}
\usepackage{rotating}

\usepackage{url}
\usepackage{listings}
\usepackage{bm}

\usepackage{float}

\usepackage[ruled, lined, longend]{algorithm2e} 

\usepackage{rotating}

\usepackage[dvipsnames]{xcolor}
\newcommand{\red}[1]{{\color{black}#1}}
\newcommand{\blueJA}[1]{{\color{black}#1}}

\usepackage{graphicx}

\usepackage{gensymb}

\usepackage[displaymath]{lineno}

\begin{document}

\renewcommand{\floatpagefraction}{.8}%
\renewcommand{\topfraction}{.75}


\title{Global optimization for data assimilation in landslide tsunamis models}

\author[1]{A.M. Ferreiro-Ferreiro }
\author[1]{J.A. Garc\'ia-Rodr\'iguez\corref{cor1} }
\ead{jose.garcia.rodriguez@udc.es}
\cortext[cor1]{Corresponding author}
\author[1]{J.G. L\'opez-Salas}
\author[2]{\\C. Escalante}
\author[2]{M.J. Castro}


\address[1]{Department of Mathematics, Faculty of Informatics, Campus Elvi\~na s/n, 15071-A Coru\~na, Spain}
\address[2]{Department of An\'alisis Matem\'atico, University of M\'alaga}

\begin{abstract}
The goal of this article is to make automatic data assimilation for a
landslide tsunami model, given by the coupling between a
non-hydrostatic multi-layer shallow-water and
a Savage-Hutter granular landslide model for submarine avalanches. The coupled model is discretized using a positivity preserving second-order path-conservative finite volume scheme. The data assimilation problem
is posed in a global optimization framework and
we develop  and compare \blueJA{parallel} metaheuristic stochastic global optimization algorithms,
more precisely  multi-path versions of the Simulated Annealing algorithm,
with hybrid global optimization algorithms based on hybridizing Simulated Annealing with 
gradient local searchers, like L-BGFS-B.
\end{abstract}


\maketitle

\section{Introduction}

\red{
The goal of this work is twofold. On the one hand, assessing the feasibility of performing data assimilation for models of tsunamis generated by submarine landslides (\blueJA{also known as} submarine mass failures, SMF), when only information/data of the fluid free surface  is available: that is, checking whether the data assimilation problem is well posed, \blueJA{i.e. the  identifiability  of the model parameters}.  On the other hand, if the former is possible, we also aim at developing a generic data assimilation framework/machinery based on parallel and efficient global optimization algorithms which can deal with landslide tsunami models.
}

\newpage

The tsunami hazard modeling is of great importance to
prevent and forecast the consequences 
of such events, as they  can cause a large number of casualties and 
huge financial losses.  Tsunamis can be generated mainly by earthquakes, storm surges or landslides (subaerial or submarine). The majority of them are caused by an offshore earthquake that pushes the ocean up or down.  Nevertheless tsunamis can also be generated in other ways. Underwater landslides, which might accompany an earthquake or occur independently, are a classic example.  Traditional warning systems completely miss tsunamis from those types of sources. Once we have a model for these phenomena, the correct calibration of the parameters is of key importance for the accurate simulation of the  tsunami. 
This calibration could be even done in real time, feeding the model 
with the measures given by the tide-gauges in the ocean, in the first moments of the 
tsunami. After the calibration, the data can be used to rerun the model and 
predict the trajectory of the tsunami and the impact areas.
 
Several types of models can be found in the literature for modeling landslide tsunamis. Their development  focuses in three aspects:  a physical model for landslide material,  a hydrodynamic model that simulates the generation and propagation of resulting waves, and the coupling between both. The hydrodynamics of landslide-induced tsunamis has been extensively studied using numerical models based on different levels of simplification.

The simplest model contemplates the landslide as a rigid solid with fixed landslide shape (see for example \cite{Grilli2005}). Another approach to simulate landslide-induced tsunamis is to consider both the landslide and the water as two different fluids (see  \cite{Fine-05, Skvortsov-07, Abadie-12,Horrillo-13, AssierRzadkiewicz97, MA-13}). This approach allows the landslide to deform, and to couple the landslide and the fluid. Although  the two-fluid models described above can be reasonably successful in predicting tsunami wave generation, they may fail to determine the landslide motion from initiation to deposition. 

Initial steps towards development of granular flow-based models for landslide behavior have usually been based on depth-integrated models pioneered by Iverson (1997, see \cite{Iverson-97}), Savage and Hutter (1989, see \cite{savage_hutter_1989}), and others. These models were initially developed for application to shallow subaerial debris flows. In \cite{FERNANDEZNIETO2008} a two-layer Savage-Hutter type model was proposed to simulate submarine landslides, where the hydrostatic pressure assumption is assumed to derive the model.

In \cite{MA201540} a two-phase model for granular landslide motion and tsunami wave generation is developed. The granular phase is modeled by a standard Savage-Hutter type model governed by Coulomb friction and the tsunami wave generation is simulated using a three-dimensional non-hydrostatic wave model, which is capable of capturing wave dispersion efficiently using a small number of discretized vertical layers. 

Here, we follow a similar approach, that is, we consider a two-phase model, however we will replace  the three-dimensional non-hydrostatic model by the multi-layer non-hydrostatic model recently proposed in \cite{Fernandez-Nieto-16}.  We briefly describe this model in Section 2.

The previous model depends on a set of parameters that  need to be calibrated
in order to match real data.  Note that, having a good model and a strong and reliable numerical method for solving the problem, is as important as performing a good parameters adjustment of the model according to physical measures. \blueJA{In other words,} a good model, together with a good numerical method, can lead to  totally wrong results with poorly calibrated parameters. Data assimilation is the tool for embedding reality in  numerical simulation. Together with mathematical modeling and development of the proper numerical methods, it could be considered as the third leg supporting 
the numerical simulation of processes in science and engineering, allowing the
model to learn and profit from real measured data, see the pioneering 
work of J. Lions about the mathematical basis of data assimilation
and control, \cite{Lions-71}. Data assimilation is of \blueJA{key} importance, for example, in atmospheric models for weather forecasting, see \cite{Kalnay-03}.

\red{
Our work follows the classical approach to calibrate the parameters of a model, i.e. the parameters are adjusted in such a way that the behaviour of the model approximates, as closely and consistently as possible, the observed response of a hydrologic system over some historical period of time. Ultimately, the best parameters are those minimizing the simple least square objective function of the residuals, which \blueJA{accounts} for the differences between the model-simulated output and the measured data. This is the right approach as long as the mathematical model is correct (realistic enough), and physical data are measured without error. The uncertainty in the model prediction will be due to the uncertainty in the parameter estimates. 

There is a separate line of research \cite{Vrugt} arguing that models have structural errors arising from the aggregation of spatially distributed real-world processes into mathematical models. Besides, due to this aggregation process, model parameters usually do not represent directly measurable entities and must therefore be estimated using measurements of the system inputs and outputs, thus adding another source of errors. As a consequence, during the calibration process one should also take also into account input, output and model structural errors. 
Several methods were firstly proposed to deal with model structural and data errors, like the Bayesian approach, Recursive Parameter Estimation algorithms, multiobjective calibration or stochastic input error models. Bayesian methods treats model parameters as probabilistic variables, in contrast with Frequentists approaches which consider model parameters fixed but unknown. Examples of Bayesian methods in hydrology are the Generalized Likelihood Uncertainty Estimation framework of Beven and Binley \cite{BevenBinley92} and the Bayesian Recursive Estimation approach of Thiemann \cite{Thiemann2001}. Recursive Parameter estimation algorithms help to identify model structural flaws by reducing the temporal aggregation associated with traditional batch processing, like PIMLI and recursive Shuffled Complex Evolution Metropolis algorithms (SCEM-UA) \cite{Vrugt2002, Vrugt2003c}. Multiobjective frameworks in order to better understand the limitation of the models, use complementary criteria in the optimization procedure and analyze the trade off in the fitting of these criteria; MOCOM \cite{Yapo} and MOSCEM-UA \cite{Vrugt} being examples of these algorithms. Finally, realistic stochastic input error models, like the Bayesian Total Error Analysis of Kavetski, only account for input errors.

These previously discussed methods were not \blueJA{successful} to account for all the referred sources of uncertainty in hydrologic modelling, i.e. parameter, input, output and structural model errors. Later, sequential data assimilation (SDA) techniques, represented by Kalman and extended Kalman filters techniques, for linear and nonlinear models respectively, continuously update the parameters of the model when new measurements are available, in order to improve the model forecast and evaluate the forecast accuracy. Recently, Vrugt et al. in \cite{Vrugt} enrich the classical calibration approach with SDA techniques, thus developing the called simultaneous parameter optimization and data assimilation (SODA) method, which combines the strengths of the parameter search efficiency and explorative capabilities of the Shuffled Complex Evolution Metropolis algorithm \cite{Vrugt2003a}, with the power and computational efficiency of the ensemble Kalman filter, thus accounting for the parameter, input, output and model structural uncertainties in hydrologic modeling.


Another approach aiming to reduce the uncertainty of models and improve their prediction skills consists on identifying the sensitive parameters and then focuss on reducing the error of these delicate parameters \cite{Yin2014}. For example, in \cite{Yuan2019}, Yuan Shijin et al. studied the sensitivity of wind stress, the viscosity coefficient and the lateral friction for the simulation of the double-gyre variation in the Regional Ocean Modeling System \cite{Haidvogel2000}, a model that can be used to simulate global waters of any size from basins to oceans. This sensitivity study was carried out not only for single parameters, but also for the combination of multiple parameters, by means of solving the Conditional Nonlinear Optimal Perturbation related to Parameter (CNOP-P) method \cite{Mu2010}, with the help of a modified Simulated Annealing (SA) algorithm in order to find the optimal solution in an efficient way.
These works (\cite{Yuan2019}) exploring  optimal parameters using sensitivity experiments, not only for individual parameters but also taking into account the interdependence between model parameters, are not feasible for models with large number of parameters, due to the fact that the number of necessary experiments increases exponentially with the involved number of model variables. A study of the sensitivities of the parameters for a symplified version of the model we are considering in this work was carried out by means of Multi-Level Monte Carlo in \cite{SANCHEZLINARES20157211}, the fluid model component being  hydrostatic with just one fluid layer.

}

In a general setting, the data assimilation problem, for a given model, can be posed as an unconstrained global optimization problem in a bounded domain.
Stochastic global metaheuristic algorithms are useful to solve these kind of problems. They have the advantage of  
needing little information of the function, and also allow to escape from local optima, being their main disadvantage
the slow rate of convergence, which is typical of Monte Carlo  algorithms. Classical well known examples of these 
methods are   Simulated Annealing (see \cite{Kirk1983,Aarts1985}), Particle Swarm (PS, see \cite{Vaz2007,Vaz2009}) or Differential Evolution (DE, see \cite{storn-price-97}). 
Conversely, local optimization algorithms are deterministic and use more information
of the function, thus being faster. Their main disadvantages are 
that, in general, they require some regularity of the cost function, and even more important, they do not
guarantee reaching the global optimimum, as they can get trapped into a local minimum.
They can be gradient free, for example Pattern Search (PS, see \cite{Hooke:1961:DSS:321062.321069}) or Nelder-Mead (NM, see \cite{nelder-mead-65}); or gradient based,
like steepest descent, Newton method, Conjugate Gradient (CG), Nonlinear CG (NCG, see \cite{polak-ribiere-69})
or Quasi-Newton methods, for example, BFGS \cite{broyden70, fletcher70, goldfarb70, shanno70}, L-BFGS \cite{liu-nocedal-89} or L-BFGS-B
\cite{byrd-lu-nocedal-zhu-95}.
One idea to profit from the good properties of stochastic (global) and deterministic (local) 
algorithms, is to hybridize them: this can be done, for example, by nesting the local search inside the global algorithm. One example is the Basin Hopping (BH) algorithm \red{\cite{Robertson89,Navon90,wales-doye-97}}. In this work, in order to calibrate the tsunami model, we follow this idea, using a in-house developed parallel multi-path version of the BH algorithm.

Data assimilation for shallow-water models has been  addressed  in many works. In these works usually gradient based local optimization methods, like the   simplest steepest descent method, have been  used to solve the  resulting optimization problem. Due to the high computational cost, the gradient is computed by solving the adjoint problem, either by  solving directly the adjoint system or computing the adjoint by automatic differentiation (AD, see \cite{Wengert1964,tapenade}).  For example, in \cite{DingYan04} the identification of Manning's roughness coefficients in shallow-water flows is performed, and the authors compare three local optimization  algorithms, a n-trust region method, L-BFGS and L-BFGS-B minimizers, where the gradients are computed by solving the  adjoint equations.
In \cite{BELANGER2005114} the variational data assimilation method (4D-VAR) is presented as a tool to forecast floods, in the case of purely hydrological flows: the cost function is a modification of the shallow-water equations to include a simplified sediment transport model and  the steepest descent algorithm is then used to find the minimum.  The initial and boundary conditions are calibrated. The gradient of the cost function is analytically computed by solving the adjoint equations of the model. 
In \cite{LAI20091} the authors developed a  4D-VAR combining remote sensing data (spatially distributed water levels extracted from spatial images, SAR) and a 2D shallow-water model to identify time-independent parameters (e.g. Manning coefficients and initial conditions) and time-dependent parameters (e.g. inflow).
In \cite{HOSTACHE2010257} the authors show the application of the technology developed in \cite{LAI20091} to derive water levels with precision from satellite images of a real event. 
In  \cite{Honnorat2009} the authors presented a method to use Lagrangian data along with classical Eulerian observations,
in a variational data assimilation process for a river hydraulics  2D shallow-water model, 
using the trajectories of particles advected by the flow  and extracted from video images. In all the cited works AD is applied for computing the gradients, and the data assimilation is performed using gradient local optimization  algorithms.

Data assimilation for tsunamis forecasting  and early warning is a very challenging problem, and on top of that  some data are even unknown, for example the geometry of the landslide or bottom deformation related to 
earthquake. Real time data is available in the Tsunami Early Warning Systems (TEWS), for example in the tide-gauges network of \blueJA{Deep-Ocean} Assessment and Reporting of Tsunamis (DART)  from National Data Buoy Center of the NOAA, or similar systems from other countries, see \cite{Bernard-15}.
Tsunami buoys are not only intended to display the occurrence of the tsunami, but also to provide real time data that can be assimilated into the tsunami warning system, to improve the accuracy of the tsunami forecasting.
Real time data assimilation in tsunamis models is mostly done using
optimal interpolation (OI) and tsunami Green functions, which are calculated in advance with linear tsunami propagation models, see for example \cite{Wang2018,Yuchen-17}. Another alternative assimilation method, is to use Kalman filter techniques (see \cite{LI20083574,NARAYAN20126401}) for wave field reconstructions and forecasts, see \cite{doi:10.1029/2018GL080644,TAKAGI-98}.
In \cite{doi:10.1029/2018GL080930}  data assimilation is done using a OI algorithm to both the real observations and virtual stations, in order to construct a complete wave front of tsunami propagation.
In \cite{Gusman-2016} tsunami data assimilation of high-density offshore pressure gauges is performed.
In
\cite{doi:10.1029/2018GL080644} a Kalman filter technique is proposed and
compared with OI. In \cite{Nodet_2006} the assimilation of Lagrangian data into a primitive equations circulation model of the ocean at basin scale, using the four-dimensional variational technique and the adjoint method, is performed.
In \cite{Tsushima2011} retrospectively data assimilation is applied to the tsunami generated in $2011$ off the Pacific coast by the Tohoku Earthquake (Mw $9.0$).
The data assimilation is done using an algorithm of near-field tsunami forecasting with tsunami data  recorded at various offshore tsunami stations: these measures were taken between 5 and 10 minutes before the tsunami reached the coastal tide-gauge stations nearest to its origin.

Nevertheless data assimilation 
in landslide generated tsunamis is not so well-developed. In this work we propose to use global optimization algorithms, that in general produce better results than the local ones. In fact  many times  the calibrated parameters do not correspond to the global minimum of the involved cost function because the considered local optimizer got stuck in a local minima far from the global solution.

\red{

Our work lies in the same vein of the  recent \blueJA{works} of Sumata et al. \blueJA{\cite{Sumata2013} and} \cite{Sumata}\blueJA{. For example in \cite{Sumata} the authors} applied a global minimization algorithm in order to calibrate an Artic Sea Ice-Ocean model. Their approach \blueJA{consists} on minimizing a cost function \blueJA{corresponding to} the model-observation misfit of three sea ice quantities \blueJA{(the sea ice concentration, drift and thickness)}, with a genetic algorithm.  The similarities between this work  and our approach are the use of bound constrained global stochastic minimization and the method to assess on the optimality of the achieved solution by using a pool of independent and randomly initialized minimization experiments. Nevertheless, the approach we are proposing differs from their strategy in several features. First of all, our goal is to calibrate a tsunami model involving less parameters than the 15 model variables of the sea ice-ocean model calibrated in their article. Besides, the different nature between this model and the tsunami model we are looking at, enforces a different optimization window, a large one (two decades) in their work versus a small one (a few hours at most) in our sketch.  \blueJA{On top of that, Sumata et al. performed the optimization of the cost function on a discrete search space, while our approach, allowing a continuous parameter domain, is richer.
}

Based on their previous work \cite{Sumata2013}, Sumata et al. in \cite{Sumata} support, as our work does, the statement that gradient descent local minimization algorithms are likely to get stuck at local minima for these complicated cost functions. Therefore, the authors impose the need to use stochastic global minimization algorithms. In fact, in  \cite{Sumata2013} two types of optimization methods were applied to the calibration of a coupled  ocean-sea ice model, and a comparison was made to assess the applicability and efficiency of  both methods. One was a gradient descent method based on finite differences for computing the gradient, while the other was a genetic algorithm. Also a parallel implementation was carried out to speed up the optimization process. In the case of the gradient descent method, each component of the gradient was computed in parallel. 
They precisely conclude that the  global optimization GA is preferred, because it yields a better optimum, since the gradient local optimizers could get trapped in local optima, even if several launches of the gradient algorithm are launched\blueJA{, in a multistart fashion}. This statement exactly coincide with our forthcoming conclusions in Section 4.1 and 4.2 (see Figures \ref{fig:test_local1_analytic} and \ref{fig:test_local_analytic}).

In our paper, we overcome this disadvantage, by proposing for first time in this field, the use of a parallel  hybrid local-global minimization algorithm.
More precisely we develop a  BH like algorithm.
BH \blueJA{consists} on hybridizing SA and local gradient searchers, allowing  to benefit from both worlds, the global convergence properties of SA and the speed of local optimizers. We go even further by proposing a parallel version of the BH algorithm. For the local searcher  ingredient of BH, we use a bounded version of the L-BFGS algorithm used in \cite{Sumata2013}, namely the L-BFGS-B algorithm. 
This version is able to increase the convergence speed and the success rate of BH.
The multistart technique performed in \cite{Sumata2013} can be seen as computing only one temperature stage of our multi-path BH algorithm.
Another advantage of our algorithm  is its embarrassingly parallel nature, as we can map each search path to a different parallel thread. In \cite{Sumata2013} each CPU thread computes one component of the gradient, while in our case, each  thread is responsible of one L-BFGS path. 
We show using an analytical test, that this algorithm improves the multi-start technique, as it is always able to find the global optimum.
Besides, in our article not only we compare the efficiency of this multi-path BH, with the equivalent version of a multipath SA (that can be seen as the BH without performing the local searches), but also show that by using the gradient searches the convergence speed of even a multipath SA increased. As mentioned  before, a SA algorithm was also used in \cite{Yuan2019} to effectively solve  the CNOP-P of ROMS.

}

The organization of this paper is as follows. In Section 2 we pose the data assimilation problem. In  Section 2.1 we describe the cost function, which is given by the measure  of the mismatch between the free surface laboratory data and the computed one, that depends on the parameters we want to assimilate. The optimization of this cost function is a hard problem: on the one hand, the evaluation of  the cost function is an expensive computational problem, because it relies in the  solution of a time dependent system of partial
differential equations.  On the other hand, this data assimilation problem gives rise to  a global optimization problem. 
In Section \ref{sec:model} we briefly describe the two-phase tsunami model and give some references about the numerical scheme we use.   The physical parameters 
of the system, that need to be calibrated, are the ratio of densities between the grain and the fluid, the Coulomb friction angle and the Manning friction coefficient. The evaluation of the cost function requires a numerical solution of this two-phase model, computed for a given set of parameters.

In Section 3, we recall the global optimization algorithms that we will use:  
multi-path Simulated Annealing  and  multi-path Basin Hopping   algorithms. Both algorithms were 
proposed by the authors in \cite{Ferreiro2013}  and \cite{Luis-BH-18} for accelerating the convergence 
of SA and BH respectively, and are based in performing synchronized parallel Metropolis 
searches, or parallel gradient based local searches. They were assessed against 
the hard benchmarks in the global optimization field, and have been successfully 
applied to \blueJA{the} calibration of models in finance, even in
the case where the costly Monte Carlo method is the only
alternative to price the involved financial products
(see for example \cite{FERREIRO2014}). In this work we apply these algorithms for data assimilation in landslide tsunami modeling. 
One of the objectives of this article is to show that this type of algorithms \blueJA{can} be successfully  applied for the parameters calibration on challenging geophysical problems.

In Section \ref{sec:numerical-results}, we present the numerical experiments that we have carried out: Section \ref{sec:analytical} is devoted to  validating the methodology using synthetic tests, in which  the model is run for fixed  sets of parameters, and we generate files with the free surface information. Then, we consider these data as data coming from laboratory, and try to recover the  parameters that were used to generate those data, by global optimization  in a large domain. After validating the methodology,  in Section \ref{sec:experimental} we apply the technique for performing the data assimilation considering  real laboratory data.

\section{Data assimilation problem}

In general, the cost function measures  the error, computed in some norm,
between the real data  and the solution produced by the numerical model.
The model will depend on a set of parameters. For example,
in the case of a one layer shallow-water model, they can be: 
one Manning coefficient for the whole domain, or also
several Manning coefficients, one per subdomain; the initial conditions; the boundary conditions, etc.
These parameters can  be even time dependant (boundary conditions, for example).

\subsection{Cost function}

In this study, the cost function only depends on the free surface elevation because this quantity is easily
measurable and  perhaps the most important magnitude to predict the tsunami inundation. 
Thus, to carry out the data assimilation method we can introduce
the following cost function using  the 
Hilbert space $L^2(0,T; \Omega)$ norm:
\begin{equation}
f(\pmb p)=\|\eta^{\pmb p}-\eta^{obs}\|_{L^2(0,T;\Omega)}= \Bigl(\int_0^T \| \eta^{\pmb p}(\cdot, t)-\eta^{obs}(\cdot,t)\|_{L^2(\Omega)}^2 dt \Bigr)^{1/2},
\end{equation}
where $\Omega\subset \mathbb{R}$ is the spatial domain, $[0,T]$ is the time domain, $\eta^{\pmb p}(x,t)$ is the free surface elevation at the point $x$ and at time $t$ computed with some model using the set of parameters $\pmb p$, and $\eta^{obs}$ are the observed values, that
can be obtained from SAR images, sea buoys or laboratory experiments.
This leads to an unconstrained global optimization \blueJA{problem} in a bounded domain. More precisely, we address problems that can be formulated as
$$
\displaystyle
\label{eq:UncOptProb}
\min_{\pmb{p} \in D \subseteq \mathbb{R}^n} f(\pmb p),
$$
where $f$ is a real valued function, with ${\pmb p}\in \mathbb R^n$  the vector of parameters,
defined on 
$ D=\prod_{i=1}^{n} \, [l_i,u_i]$, with $l_i$
and $u_i$ being the lower and upper bounds in direction $i$, respectively.
The solution can be written as:
$$
\pmb p^*=\underset{\pmb{p} \in D \subseteq \mathbb{R}^n}
{\text{arg min}}  f(\pmb p).
$$

In the discrete case, the cost function will have the following expression:
$$
f({\pmb p})=\sqrt{\sum_{k=1}^{N_T} \sum_{i=1}^{N} 
\left(\eta^{\pmb p}_{i,k} -\eta^{obs}_{i,k}\right)^2},
$$
where
$\eta_{i,k}^{\pmb p}=\eta^{\pmb p}(x_i,t_k)$  and  $\eta_{i,k}^{obs}=\eta^{obs}(x_i,t_k)$ being
$ x_i$  the $i$-th measure point, for $i=1, \ldots, N$ and
$ t_k$  the $k$-th measure time, with $k=1,\ldots, N_T$.

\begin{figure}[!htb]
\begin{center}
\includegraphics[width=12cm]{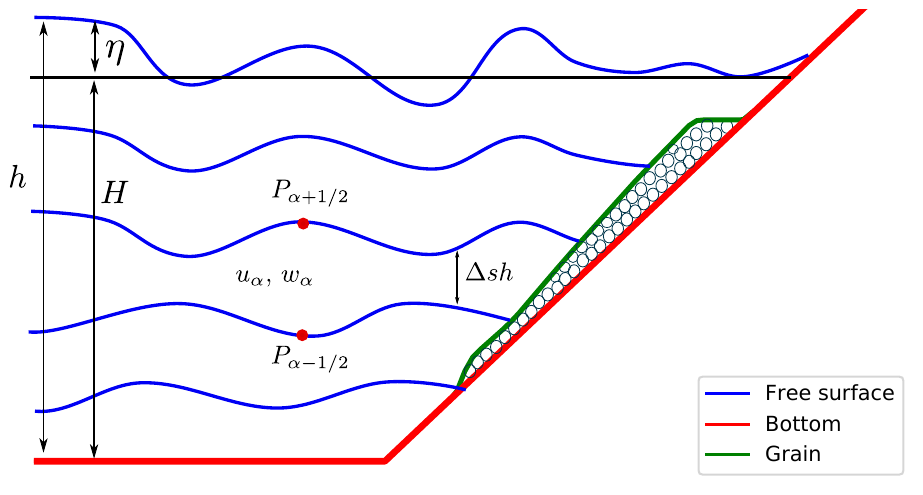}
\end{center}
\caption{Sketch of the model.}
\label{fig:channel-sketch-multi-layer}
\end{figure}

Note that the cost function depends on $\eta^{\pmb p}$\blueJA{,} which implicitly depends on the parameters to be calibrated. Therefore, a single evaluation of the cost function requires a realization of the numerical model for a given set of parameters. In the next section we present the equations of the two-phase model, pointing out what are the parameters to be calibrated. Some basic idea about the numerical scheme we use is also sketched.

\subsection{Mathematical model \label{sec:model}}

As discussed in the introduction, we use a two-phase model in order to describe the interaction between the submarine landslide and the fluid.  In this work,  a Savage-Hutter  model  (see \cite{savage_hutter_1989}) is considered for the kinematics of the submarine landslide, and a multi-layer non-hydrostatic shallow-water model is used for the evolution of the ambient water (see \cite{Fernandez-Nieto-16}). Both models are coupled through the boundary conditions at the sea-floor.

At this point, we suppose that the landslide is totally submerged and that the ratio of densities between the ambient fluid and the granular material is given by the parameter $r$. Usually
\[
r=\frac{\rho_f}{(1-\varphi) \rho_s + \varphi \rho_f}\, ,
\]
where $\rho_s$ is the typical density of the granular material, $\rho_f$ is the density of the fluid ($\rho_s> \rho_f$), and $\varphi$ is the porosity ($0\leq \varphi <1$). Here, we suppose that $\varphi$ is constant on space and time, and therefore $r$ is also constant. Note that $0<r<1$. Finally, let us remark that even on a uniform material, $r$ is difficult to estimate as  it depends on porosity $\varphi$. Typical values of $r$ are in the interval  $[0.3,0.8]$.

The 1D Savage-Hutter model that we consider in this article is written as follows:

\begin{equation}\label{SH}
 \left\{
 \begin{array}{l}
\displaystyle {\partial_t z_s +\partial_x(z_su_s)}= 0,  \\[0.3cm]
\displaystyle{\partial_t (z_su_s) +\partial_x\left( z_su_s^2+\frac{g(1-r)}{2}z^2_s\right)} =g(1-r)z_s\partial_x H + \tau_C, \\
 \end{array}
  \right.
\end{equation}
where $g$ is the gravity acceleration ($g=9.81$ $m/s^2$); $H(x)$ is \blueJA{the} non-erodible bathymetry measured from a given reference level and unchanged during the simulation; $z_s(x,t)$ is the landslide depth at each point $x$ at time $t$; and $u_s(x,t)$ the averaged horizontal velocity.  $\tau_C$ is the Coulomb friction term given by:
\[
\tau_C= - g (1-r) \mu z_s \frac{\sqrt{u_s^2}}{u_s}.
\]
Note that this term is multi-valuated when $u_s=0$. The simplest friction law corresponds to a constant friction coefficient:
\begin{equation} \label{def_mu}
\mu=\mbox{tan}(\theta),
\end{equation}
where $\theta$ is the friction angle, although more complex friction terms could  be used to simulate natural subaerial or submarine landslides (see \cite{mangeney_2007,Pirulli2008}).  Other definitions, derived from experimental data, have been proposed by Pouliquen (see \cite{pouliquen99}) where the friction coefficient depends on the velocity and thickness of the granular layer. This law is widely used in the literature and involves at least three parameters to be calibrated (see e.g. \cite{Brunet2017}).

The Coulomb friction term $\tau_C$ is quite relevant, as it controls the motion of the landslide. In particular, it is defined in terms of the friction angle $\theta$, 
which is a parameter to calibrate in order to fit the simulation with the experimental data. Finally, let us mention that in the derivation of the previous model we have supposed a rigid-lid assumption with respect to the free surface of the ambient fluid: that is, the pressure variations induced by the fluctuation on the free surface of the ambient fluid over the landslide  are neglected. Nevertheless, the buoyancy effects have been taken into account.

The ambient fluid is supposed to be modeled by a multi-layer non-hydrostatic shallow-water system recently proposed in \cite{Fernandez-Nieto-16}. This system is obtained
by a process of depth-averaging of the incompressible Euler equations. More precisely, it can be seen as a particular semi-discretization with respect to the vertical variable
of the incompressible Euler equations. Total pressure is decomposed into a sum of a hydrostatic  and a non-hydrostatic component. In this process, vertical velocities are assumed to have a linear  vertical profile, whilst the horizontal velocities are supposed to have a constant vertical profile. The resulting  multi-layer model admits an exact energy balance, and when the number of layers increases, the linear  dispersion relation of the
linear model converges to the same of Airy's theory. The model proposed in 
\cite{Fernandez-Nieto-16} can be written in compact form as

\begin{equation}
\left\{
\begin{array}{l}
\partial_t h + \partial_x (h\bar u) = 0,\\[0.3cm]
\\
 \displaystyle{\partial_t (hu_\alpha ) + \partial_x \Bigl(hu_\alpha^2 + \frac{g}{2}h^2\Bigr) - 
gh\partial_x(H-z_s)}  \\[0.3cm]
\qquad \qquad +u_{\alpha +1/2}\Gamma_{\alpha +1/2} - u_{\alpha -1/2}  \Gamma_{\alpha-1/2} = - h (\partial_x p_\alpha  + \sigma_\alpha  \partial_z p_\alpha  ) - \tau_\alpha,\\[0.3cm]
 \displaystyle{\partial_t (hw_\alpha ) + \partial_x (hu_\alpha  w_\alpha  ) + w_{\alpha +1/2} 
\Gamma_{\alpha +1/2} - w_{\alpha-1/2} \Gamma_{\alpha -1/2}} = - h\partial_z 
p_\alpha,\\[0.3cm]
 \partial_x u_{\alpha -1/2} + \sigma_{\alpha -1/2} \partial_z 
u_{\alpha -1/2} + \partial_z w_{\alpha -1/2} = 0,
\end{array}
\right.
\label{eq:system-multi-layer}
\end{equation}
for $\alpha\in \{1,2,\ldots,L\}$, being $L$ the number of layers. In the previous system, we have used the following notation:

\begin{equation}
\begin{array}{c}
u_{\alpha+1/2} =
 \dfrac{1}{2}(u_{\alpha+1} + u_{\alpha} ) ,\quad \partial_z u_{\alpha+1/2} =
\dfrac{1}{h\Delta s} (u_{\alpha+1} - u_{\alpha} ) ,

 \\
 \\
w_{\alpha+1/2} =
\dfrac{1}{2} (w_{\alpha+1} + w_{\alpha} ) ,\quad \partial_z w_{\alpha+1/2} =
\dfrac{1}{h\Delta s} (w_{\alpha+1} - w_{\alpha} ),
\\
\\
p_{\alpha} =
 \dfrac{1}{2}(p_{\alpha+1/2} + p_{\alpha-1/2 }) ,\quad \partial_z p_{\alpha} =
 \dfrac{1}{h\Delta s}(p_{\alpha+1/2} - p_{\alpha-1/2 }),
 \\
 \\
\sigma_{\alpha}= \partial_x (H-z_s - h\Delta s(\alpha - 1/2)) ,\quad \sigma_{\alpha-1/2} = \partial_x (H-z_s - h\Delta s(\alpha - 1)).
\end{array}
\end{equation}
As depicted in Figure \ref{fig:channel-sketch-multi-layer}, the flow depth $h$ is split along the vertical axis into $L \geq 1$ layers and
$\Delta s = 1/L$. $u_\alpha$ and $w_\alpha$ are the depth averaged velocities in the $x$ and $z$ directions respectively, and $g$ is the gravitational acceleration. The term $p_{\alpha+1/2}$ is the non-hydrostatic pressure at the interface $z_{\alpha+1/2}$. The free surface  elevation measured from the still-water level is $\eta = h - H+z_s$, where again $H(x)$ is the unchanged non-erodible bathymetry measured from a given reference level. $\tau_\alpha=0$, $\alpha>1$ and $\tau_1$ is the Manning friction term  that is only present at the lowest layer ($\alpha=1$) given by
\[
\tau_1= g h\frac{n^2}{h^{4/3}} u_1|u_1|.
\]
Finally, for $\alpha=1, \ldots, L-1$, $\Gamma_{\alpha+1/2}$  account for the mass transfer across interfaces and   are defined by
\[
 \Gamma_{\alpha+1/2}=\sum_{\beta=\alpha+1}^L
 \partial_x (h\Delta s (u_\beta  - \bar{u})), \quad  \bar{u} =
 \sum_{\alpha=1}^L \Delta s u_\alpha.
\]
Here we suppose that $\Gamma_{1/2}=\Gamma_{L+1/2}=0$, that is, there is no mass transfer through the bottom nor the free-surface.

In order to close the system, the following boundary conditions are con\-si\-de\-red:
$p_{L+1/2} = 0$, $u_0 = 0$ and $w_0 = \partial_t z_s$. Note that the last two conditions enter into the incompressibility condition for the lowest layer $(\alpha=1)$, given by
\[
 \partial_x u_{1/2} + \sigma_{1/2} \partial_z 
u_{1/2} + \partial_z w_{1/2} = 0.
\]
Observe that both models are coupled through the unknown $z_s$, present in the equations and in the boundary condition ($w_0=\partial_t z_s$).

Note that the two-phase model depends on three
coefficients (that are the ones  to be calibrated), namely the vector of coefficient is  $\pmb p=(r,\theta, n)$,
where $r$ is the ratio of densities between the fluid and the granular phase, $\theta$
the Coulomb friction angle, and $n$ the friction (Manning) coefficient. In particular the  first two are quite relevant for the landslide motion and therefore, for the induced tsunami water waves.

System \eqref{SH}  could be written in the following compact way:
\begin{equation} \label{SHC}
    \partial_t U_s + \partial_x F_{s}(U_s)=G_s(U_s) \partial_x H -S_{s}(U_s),
\end{equation}
being
\[
U_s=\left[ 
\begin{array}{c}
z_s \\[0.2cm]
u_s z_s
\end{array}
\right], \ 
F_s(U_s)= 
\left[
\begin{array}{c}
z_s u_s\\[0.2cm]
\displaystyle{z_s u_s^2 +\frac{g(1-r)}{2}z_s^2}
\end{array}
\right],
\]
\[
G_{s}(U_s)=\left[
\begin{array}{c}
0 \\[0.2cm]
g(1-r) z_s
\end{array}
\right], \
S_s(U_s)= \left[
\begin{array}{c}
0 \\[0.2cm]
\tau_C
\end{array}
\right].
\]

The multi-layer non-hydrostatic shallow-water system could also be expressed in a similar way:
\begin{equation}\label{MLC}
\left\{
\begin{array}{l}
\partial_t U_f + \partial_x F_{f} (U_f) + B_{f} (U_f)\partial_x U_f = G_{f} (U)\partial_x (H-z_s)+\mathcal{T}_{NH}- S_f(U_f), \\[0.2cm]
B(U_f,(U_f)_x,H,H_x, z_s, (z_s)_x)=0, \\
\end{array}
\right.
\end{equation}
where
$$
U_f=\left[
\begin{array}{c}
h\\
 hu_1
\\
\vdots
\\
hu_L\\
hw_1\\
\vdots
\\
hw_L
 \end{array}
 \right],
F_{f} (U_f) = 
\left[
\begin{array}{c}
h\bar u\\
hu_1^2+\dfrac{1}{2} gh^2
\\
\vdots
\\
hu_L^2+\dfrac{1}{2} gh^2\\
hu_1w_1\\
\vdots
\\
hu_Lw_L
 \end{array}
\right]
,
G_{f} (U_f) = 
\left[
\begin{array}{c}
0\\
 gh
\\
\vdots
\\
gh\\
0\\
\vdots
\\
0
 \end{array}
\right].
$$
$B_{f}(U_f)\partial_x (U_f)$ contains the non-conservative products involving the momentum transfer across the interfaces
\[
B_f(U_f)\partial_x (U_f)= 
\left[
\begin{array}{c}
0 \\[0.2cm]
u_{3/2}\Gamma_{3/2} \\[0.2cm]
u_{5/3}\Gamma_{5/2} - u_{3/2} \Gamma_{3/2} \\[0.2cm]
\vdots \\[0.2cm]
- u_{L-1/2}\Gamma_{L-1/2} \\[0.2cm]
w_{3/2}\Gamma_{3/2} \\[0.2cm]
w_{5/3}\Gamma_{5/2} - w_{3/2} \Gamma_{3/2} \\[0.2cm]
\vdots \\[0.2cm]
- w_{L-1/2}\Gamma_{L-1/2} 
\end{array}
\right],
\]
$S_f(U_f)$ contains the Manning friction term
\[
S_f(U_f)= \left[
\begin{array}{c}
0 \\[0.2cm]
\tau_1 \\[0.2cm]
0 \\[0.2cm]
\vdots \\[0.2cm]
0 
\end{array}
\right].
\]
The non-hydrostatic corrections in the momentum equations are given by
\[
{\cal T}_{NH}={\cal T_{N H}}(h, h_x , H, H_x , z_s, (z_s)_x, p, p_x )= 
-\left[
\begin{array}{c}
0\\
 h(\partial_xp_1+\sigma_1\partial_z p_1)
\\
\vdots
\\
h(\partial_xp_L+\sigma_L\partial_zp_L)\\
h\partial_zp_1
\\
\vdots
\\
h\partial_zp_L
 \end{array}
\right],
\]
and finally, the operator related with the incompressibility condition at each layer is given by: 
\[
B(U_f, (U_f)_x , H, H_x, z_s, (z_s)_x ) = 
\left[
\begin{array}{c}
 \partial_x u_{1/2} + \sigma_{1/2} \partial_z u_{1/2} + \partial_z w_{1/2}
\\
\vdots
\\
\partial_x u_{L-1/2} + \sigma_{L-1/2} \partial_z u_{L-1/2} + 
\partial_z w_{L-1/2}
 \end{array}
\right].
\]

The discretization of systems \eqref{SHC} and \eqref{MLC} becomes difficult. In this article, we have considered the natural extension of the numerical schemes proposed in \cite{Escalante-17} and \cite{Escalante2018}, where a splitting  technique has been described. Firstly,  the systems \eqref{SHC} and \eqref{MLC} can be expressed as 
the following non-conservative hyperbolic system:
\begin{equation}\label{Eq:C}
\left\{
\begin{array}{l}
\partial_t U_s + \partial_x F_s(U_s)=G_s(U_s) \partial_x H, \\[0.2cm]
\partial_t U_f+ \partial_x F_f(U_f)+B_f(U_f)\partial_x(U_f)=G_f(U_f)\partial_x (H-z_s). 
\end{array}
\right.
\end{equation}
Both equations are solved simultaneously using the same {\em time step}, by means of a second order HLL,  positivity-preserving and well-balanced, path-conservative finite volume scheme (see \cite{PVM}). 
The synchronization of time steps is done taking into account the CFL condition of the complete system \eqref{Eq:C}. A first order estimation of the maximum of the wave speed for system \eqref{Eq:C} is the following:
\[
\lambda_{\max}=\max(|u_s|+\sqrt{(g(1-r) h_s}, |\bar u| + \sqrt{g h}).
\]

Next, the non-hydrostatic pressure corrections $p_{1/2}, \cdots, p_{L-1/2}$ at the vertical interfaces are computed from
\[
\left\{
\begin{array}{l}
\partial_t U_f={\mathcal T}_{NH}(h,h_x,H,H_x, z_s,(z_s)_x, p,p_x), \\[0.2cm]
B(U_f,(U_f)_x, H,H_x,z_s, (z_s)_x)=0. 
\end{array}
\right.
\]
This requires the discretization of an elliptic operator by means of standard second order central finite differences. The resulting linear system is solved using an iterative Scheduled Jacobi method (see \cite{Adsuara-16}). Finally, the horizontal and vertical momentum at each layer are updated using the computed non-hydrostatic corrections. At this stage, the frictions $S_s(U_s)$ and $S_f(U_f)$ are also discretized (see \cite{Escalante-17, Escalante2018}). We refer the reader to \cite{FERNANDEZNIETO2008} for the discretization of the Coulomb friction term.

\section{Multi-path BH global optimization \label{sec:algorithm}}

In this section we describe the 
optimization algorithms multi-path SA (SA$_{\text{M}}$) and multi-path BH (BH$_{\text{M}}$), that can be seen 
as a modification of the sequential BH algorithm, introducing a parallel multi-path searching technique.

The BH algorithm is a hybrid between the Metropolis algorithm and some
kind of gradient local optimization method, in order to profit from the speed and accuracy of the local optimizer,
while retaining the global convergence properties of the stochastic one.
\red{The seminal idea was presented by Navon and Robertson et al. in \cite{Robertson89,Navon90} 
for finding the global \blueJA{minimum} of Potential Energy Surfaces (PES) related to structures of mixed
Argon-Xenon clusters.  The authors developed the finite-temperature lattice based
Monte Carlo method and  compared the use of three different limited memory Quasi-Newton-like 
conjugate gradient methods as local minimizers, the L-BFGS against two others, being L-BFGS the better 
performing one. 
Seven years later, a similar idea was also successfully applied by
 Wales and Doye (see \cite{wales-doye-97}) in order to minimize the PES, for finding 
Lennard-Jones clusters using a nonlinear conjugate gradient  method (Polak Ribi\`ere \cite{polak-ribiere-69})
as the local optimizer. In the latter reference the authors \blueJA{named} the method  Basin-Hopping;
this name became widely accepted for referring  to these kind of global optimization methods.
Nowadays, the term BH encompasses a family of algorithms obtained by combining different local (NCG, BFGS, ...) and global stochastic algorithms (Metropolis or SA):  quasi-Newton methods (BFGS and descendants) are the most common choice for the local component.}
BH methods have been extensively studied by 
Locatelli et al., see \cite{locatelli-05, locatelli-schoen-13, addis-locatelli-schoen-05, adis-phd}, and Leary \cite{Leary}.
In the BH method, the local optimizer can be seen as an operator that transforms the original function  $f(\pmb{x})$,
returning a new piece-wise constant function, $L(\pmb{x}) = f(\mathcal{LS}(\pmb{x}))$, being
$\mathcal{LS}(\pmb{x})$ the point where a local minimum of $f$ is obtained from a starting point $\pmb{x}$.
The resulting global optimization problem for the function $L(\pmb{x})$, is much more tractable for the global optimizer component, as the barriers between local minima have been softened.

The idea of BH is to use a temperature process like in SA: we denote by
${\mathcal T}$ and ${\mathcal T}_{min}$ the current and minimum temperatures, we consider the temperature reduction schedule, ${\mathcal T}_{k-1}=\rho {\mathcal T}_{k}$, being $\rho$ the cooling rate, and we perform a Metropolis process with $N$ steps at each temperature level. More precisely, at temperature level ${\mathcal T}_k$, being $\bm{x}_k$ the starting point,
first, we generate a random neighbor, $\bm{y}_k$, inside a ball with radius ${r}_k$ and centered in $\bm{x}_k$,
$\bm{y}_k\in B(\pmb x_k,r_k)$. 
Next, we perform a gradient local  search starting from $\bm{y}_k$, in order to obtain a local minimum,  and we decide whether to accept or discard it, using the Boltzmann law. 
Finally, we advance to the next temperature level.
The algorithm stops when the temperature reaches ${\mathcal T}_{min}$, or the number of successive rejections exceeds $J$.
The radius $r_k$ is updated after a certain interval, by using the 50\% acceptance rule \cite{goffe-96}. A nice property is that BH can also be seen as a generalization of SA: SA can be recovered by skipping the local optimization phase in BH.

\begin{figure}[htbp]
\begin{center}
\includegraphics[scale=0.16,angle=90]{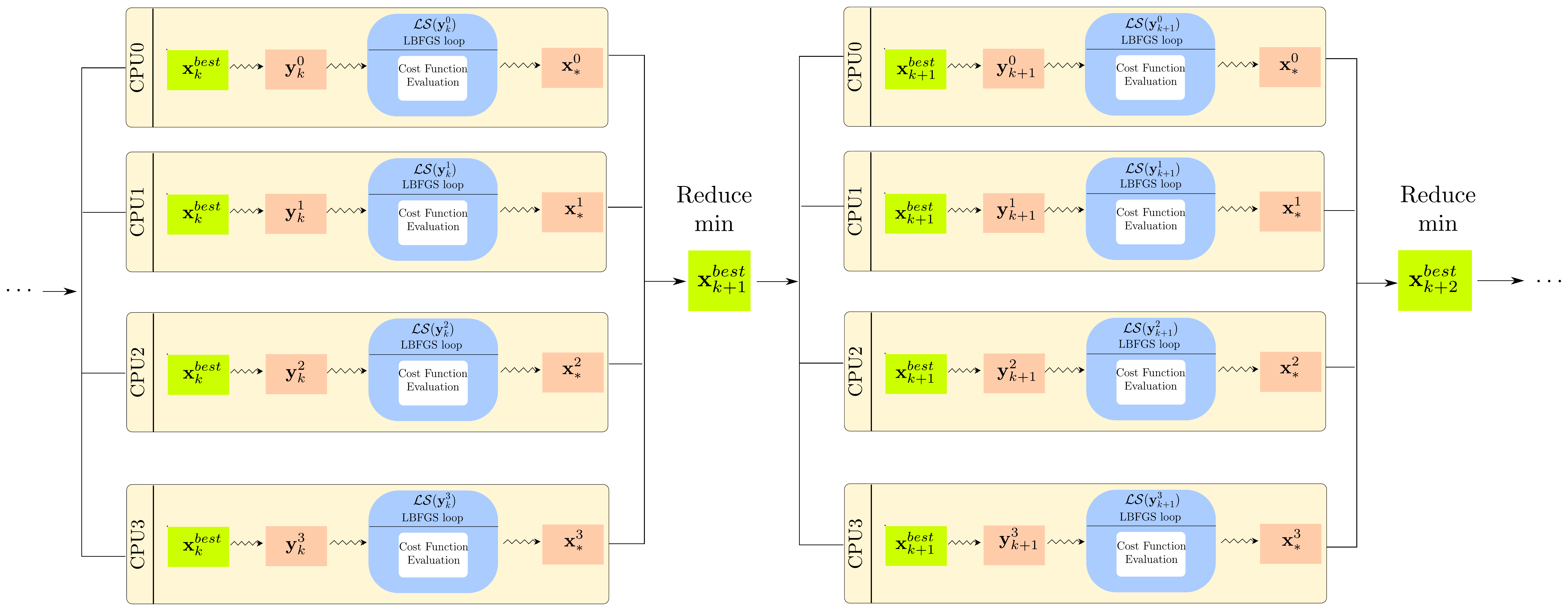}
\end{center}
\caption{Schematic visualization of the BH$\text{M}$ algorithm \blueJA{(with $\text{M}=4$)}.}
\label{fig:alg_sk}
\end{figure}

In \cite{Ferreiro2013,Luis-BH-18} the authors proposed
a synched multiple Metropolis path approach for SA and BH-like algorithms, respectively.
The idea is to perform not one, but $M$ Metropolis searches at each temperature level ${\mathcal T}_{k}$, from the same initial point $\bm{x}_k$
(see Algorithm \ref{alg:sync_BH}).
In the simplified case with $N=1$,  the  algorithm consists of 
launching $M$ gradient local searchers (see Figure 
\ref{fig:alg_sk}), starting from the corresponding set of random neighbors $\pmb y_k^l \blueJA{,\ l=1,\ldots,\text{M},}$
of the current minimum point, and thus the Metropolis searches are
entirely replaced by local searches. After performing the $N$ steps of Metropolis
at each path, and before advancing to the next temperature level, we gather the final information, keeping the best of the attained minima, so that $\bm{x}^{best}_{k+1} = \blueJA{min(\bm{x}^l_*)}$ (see Figure \ref{fig:alg_sk} \blueJA{and Algorithm  \ref{alg:sync_BH}}). We will refer to this algorithm as BH$_\text{M}$, $M$ being  the number of paths
(number of Metropolis processes with local searchers; or just the number of local searches, if $N=1$) launched at each temperature optimization step. 
Note that if besides $M=1$, then BH$_1$
corresponds with the classical BH (only one Metropolis path, or only one local search).
Also, if we replace the local search operator, $\cal{LS}$, with the identity, $id$, we recover the 
multi-path SA algorithms, SA$_\text{M}$ \cite{Ferreiro2013}; furthermore SA$_\text{1}$
corresponds to the classical single path SA.
These multi-path BH$_\text{M}$ algorithms have two interesting properties:
on the one hand, they are highly parallelizable;
on the other hand 
they improve convergence properties, both the convergence
speed and the success rate of the classical SA and BH.

\begin{algorithm}[ht!]
\caption{Synched multi L-BFGS-B BH, pseudocode.}
\label{alg:sync_BH}
$\boldsymbol{y}$ = random uniform in $D$;

Set \# successive rejections: $j = 0$\;
Iteration number: $k = 0$\;
Initial position: $\boldsymbol{x}_0 = \boldsymbol{x}_* = \mathcal{LS}(\boldsymbol{y})$\;

\While{($j < J$) \textbf{or} (${\mathcal T}<{\mathcal T}_{min}$)}{
	\For{$l=$ $1$:$M$}{
		\For{$i=$ $1$:$N$}{
			$\boldsymbol{y}_{i}^l$ = random uniform in $B(\bm{x}_{i}^l,{r}_k)$\;
			$\bm{u}\:\,$ = random uniform in $[0,1]$\;
			$\Delta\;$ = $L(\bm{y}_{i}^l)- L(\bm{x}^l_*)$\;
			\eIf{$\bm{u} < exp(-\Delta/{\mathcal T})$}{
				$\bm{x}^l_* = \bm{x}_{i+1}^l = \mathcal{LS}(\bm{y}_{i}^l)$\;
				$j = 0$\;
			}{
				$j = j + 1$\;
			}
		
		}	
			
	}
	
	Synchronization: $\bm{x}^{best}_{k+1} = min(\bm{x}^l_*)$\;
	\For{$l =$ $1$:$M$}{
		$\bm{x}^l_* = \bm{x}^{best}_{k+1}$\;
	}
	$k = k + 1$\;
	Update  ${r}_k$\;
	${\mathcal T} = \rho\cdot {\mathcal T}$\;
}

\end{algorithm}

This approach has the advantage of being easily parallelizable,
because the multiple search paths can be computed asynchronously at the same time.
For example, if we have a multi-CPU architecture, each CPU thread can 
take care of computing one local search, and after that the results have to be synchronized (see Figure  \ref{fig:alg_sk}).
In this paper, we will use this multi-path implementation in a
multi-CPU setting, each CPU thread will take care of a search path.

Regarding the convergence properties, in \cite{Luis-BH-18}
the  study of the optimal number of multi-searches, both from the convergence rate and the success rate viewpoint,
is done empirically.
According to the results, increasing the number of searchers improves the convergence rate, although this increase in
convergence rate is not unlimited.
For example, if the problem
is simple and/or the dimension is low, by
increasing the number of searchers one would only obtain a marginal increase in convergence speed.
Nevertheless, even in those cases, 
the computing time can be lower
because the evaluations can be done in parallel,
and thus this increase in the number of search paths comes almost for free.
Even if the problem is computationally hard, it always comes a point
where the optimal convergence rate is achieved and a further increase in the number of searchers will not have any advantage. 
Usually this number of searches for obtaining an optimal convergence rate is moderate: the optimization problem has to be really tough in order to demand a  high number of local searchers.
The good properties of the proposed algorithm  also apply to the success rate and the same conclusions can be obtained. Usually, it comes a point
when a $100\%$ success rate is achieved, more number of searchers will not have any advantage.
Besides, the number of searches for obtaining this $100\%$ success rate is, once more, normally moderate.
For tough problems, the 
advantage of performing a large number of local searches becomes more evident.

\red{
In this work, for the local optimizer we will use the very \blueJA{robust} L-BFGS-B algorithm.
This minimizer is intended for problems in which information on the Hessian matrix is difficult to obtain. 
It was presented by Nocedal in \cite{Nocedal97} as an extension of the L-BFGS minimizer, being a limited-memory quasi-Newton algorithm  (it does not need to store the Hessian matrix) that  allows to solve  nonlinear optimization problems
with restrictions given by simple bounds on the variables of the function to be optimized.  

In our work since the parameters are known to vary between given bounds, and we need to ensure that the optimizer
would never explode by following a wrong path outside the physical domain, we used the  L-BFGS-B  bounded gradient method. 
If one uses a non  bounded gradient local optimizer, some search paths could reach points outside the physical domain, where the equations \blueJA{could} stop making sense. In that case the evaluation of the cost function (a finite volume solver) may explode, 
either by crashing or by entering in a very low $\Delta t$ state (imposed by the CFL condition). As a consequence the assimilation process will crash or never end.
We preferred to stay safe with the bounded algorithm,
as it has almost the same computational cost as the unbounded  L-BFGS version.
}

\red{

In order to compute the \blueJA{partial} derivatives \blueJA{with respect to the variables to be identified}, needed for the gradient of the objective function, we can use either algorithms based on the so-called adjoint method or the standard finite-difference method. 
Both techniques have their own advantages and disadvantages. In this article we opted for the finite difference procedure attending to the reasons that will be discussed  hereafter.
}

\red{

There are two different approaches for tackling the adjoint problem. One technique is the classical approach developed by Lions (see \cite{Lions-71}) and applied for the  simpler 2D one layer shallow water model by Monnier et al. in \cite{LAI20091}. It consists in computing the adjoint PDE system, and then solving it by numerical methods. This is a very challenging problem even for the simpler shallow water model assimilated by Monnier, and even much more for our problem at hand: we emphasize that we are dealing with a coupled model involving an arbitrary number of fluid layers (denoted by $L$ in the PDE system \eqref{eq:system-multi-layer}\blueJA{)} of ``shallow-water type systems'', along with the Savage-Hutter equations, thus resulting in a large hyperbolic system of coupled conservation laws. The mentioned system can only be numerically approximated by means of very involved finite volume numerical discretizations, thus dealing with the corresponding stability issues related to high nonlinearities involved in hyperbolic problems along with  spatial-temporal discretization issues. As a consequence, the adjoint method will lead to a system of conservation laws with source terms and non-conservative products, for which it would not be clear the hyperpolicity region. Besides, the  numeric approximation of this adjoint system will be very sophisticated. 
One wonders if all this challenging work, \blueJA{even if feasible}, is worth it for calibrating just this particular model.
On the other hand, a way to circumvent those difficulties and avoid computing the adjoint system, is to compute the partial derivatives by means of Automatic Differentiation (AD).
As in the close future we pretend to tackle real two dimensional problems, which involve much higher computational cost, and consequently even more for the adjoint AD procedure, speeding up on GPUs the cost function evaluation (i.e. the solution of the system) becomes compulsory. In this scenario, also the automatic differentiation algorithm should be carried out  in the GPU side. Therefore, an AD library for GPUs in needed, something that can be an obstacle due to the fact that these tools are not always available, specially for massively parallel architectures like GPUs. Furthermore, the code should be rewritten from the very basics using the overloaded operators provided by the AD library. On top of that, more memory will be needed in this adjoint setting, which is again an issue in GPUs.

Having in mind all the previously discussed issues, in this article we opted for the direct numerical approximation of the partial derivatives involved in the gradient using finite differences. In our case this has several advantages when compared to the adjoint computation.
First of all, one of our goals is to  develop a data assimilation framework/machinery for landslide tsunami models, generic enough in the sense that it should be \blueJA{directly} applicable  if one wants to enrich the here considered model with further characteristics or even fully replace it with other models. 
This machinery should endowe us with a tool for comparing the accuracy of (possibly quite) different models, and this is a reason for not developing an algorithm that is too tailored/tight for a particular 
model or numerical scheme. 
In this sense, by computing the gradient via finite differences
we gain generality, since the method can be easily applied to models of all kinds without changes in the calibration procedure (in the same vein of \cite{Sumata2013}); one will just need to invoke it by plug in the new model solver (no changes are needed in the solver, unlike with the adjoint method). 
Hence we are well positioned in order to face the calibration of the previously mentioned oncoming richier two dimensional model to real data.
Additionally, our technique is able to cope with the strongly nonlinear relation between model state and parameters, for which other approaches based on Kalman filter have difficulties.
Finally, regarding the computational efficiency, finite-difference method for computing the derivatives of the cost function with respect to the parameters to be calibrated is not much more computing time demanding than the adjoint method if the number of optimization variables is short; indeed this is the case we are dealing with, our goal is to calibrate three parameters, namely the ratio $r$ of densities between the fluid and the granular phase, the Coulomb friction angle $\theta$, and the friction Manning coefficient $n$. 
{Last but not least,
nowadays, thanks to the available high computational power, the numerical computation of the gradient could be directly addressed 
making use of parallel codes that combine multi-CPU implementantion of
the optimizer and multi-GPU implementation of the numerical solver used to evaluate the cost function. 
}

}

\red{
All in all,} the gradient of the cost function will be numerically computed, using 
first order progressive finite differences
$$
\dfrac{\partial f}{\partial p_i}(\pmb p)=\dfrac{f(\pmb p+\varepsilon \pmb e_i)-f(\pmb p)}{\varepsilon},
$$
with $\varepsilon=10^{-6}$, and $\pmb e_i=(0,\ldots,1,\ldots,0)$ the 
unitary vector of direction $i$.

Regarding the implementantion of the algorithms, 
the whole
implemententation of both the  cost  function (finite volume solver) and its gradients, and the optimization algorithms\blueJA{,}
is custom made. Both algorithms have been integrated in an efficient code using C++, and
OpenMP is used for the parallel implementantion of the optimization codes (see Figure 
\ref{fig:alg_sk}).
Also we want to emphasize
 that the cost function is integrated with the optimization
tool, so that it is called on the fly for each set of parameters during the optimization
process. Therefore no intermediate results need to be discharged from RAM to the hard drive for 
computing the value of the cost function, thus resulting in an efficient code.
Furthermore,  during the
whole optimization process the laboratory data is read only once at the beginning.

\begin{figure}[!htb]
\begin{center}
\includegraphics[width=11.5cm]{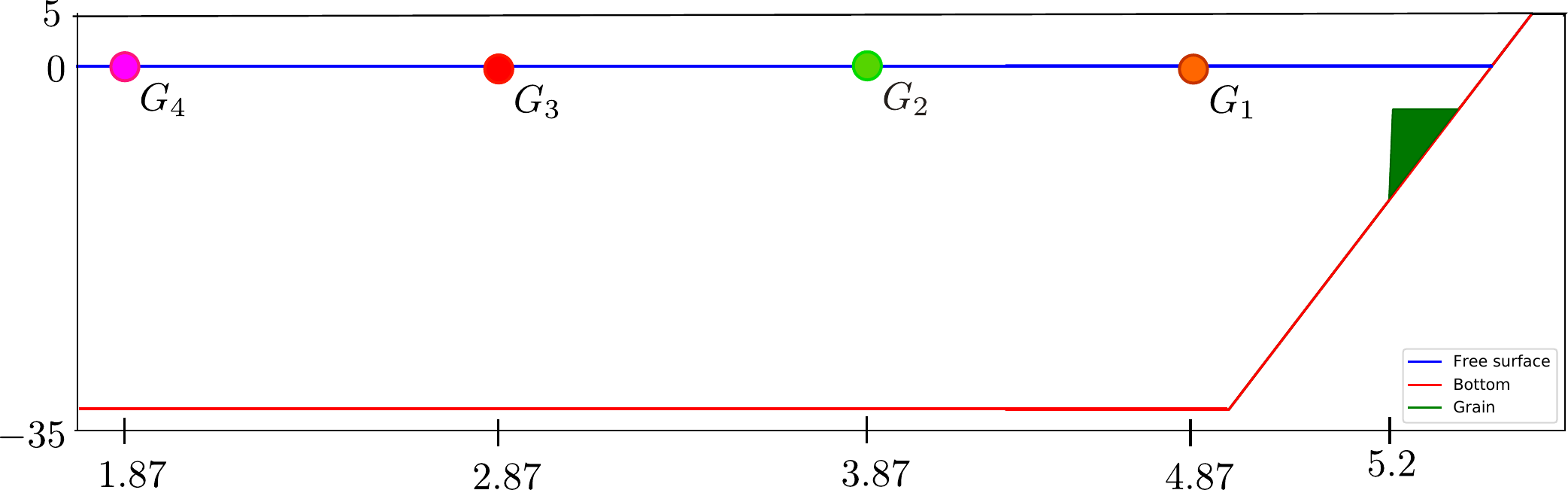}
\end{center}
\caption{Sketch of the channel, initial condition and position of the tide-gauges.}
\label{fig:channel-sketch}
\end{figure}

\section{Numerical results \label{sec:numerical-results}}
\red{
In this section we present two \blueJA{sets} of numerical examples.
The first one in Section \ref{sec:analytical} is a pool of synthetic tests
with known solutions, that \blueJA{are} used to validate 
the proposed algorithms and methodology,  to discuss about the 
identifiability of the problem, and
to show the convergence results and computational speedup.
The second one in Section \ref{sec:experimental}  shows an application of the proposed methodology to the assimilation of real laboratory data.
}

The laboratory experiment that  will be calibrated in this article was presented 
in \cite{Grilli2017}, and the data can be accessed at \cite{web-experiments}. In that work, the authors \blueJA{design different laboratory experiments and perform  numerical simulations to validate a landslide tsunami model and to asses how tsunami hazard from SMFs is affected by slide kinematics and rheology.}

In \cite{Macias2017}
the Tsunami-HySEA model is used to perform some of the numerical benchmark problems proposed \blueJA{in \cite{Grilli2017}. The obtained results are documented} in the ``Proceedings and results of the 2011 NTHMP Model Benchmarking Workshop''. 

In our article we focus in one of the experiments performed in \cite{web-experiments}: the benchmark 4 (deformable submarine \blueJA{landslide}).
For both the analytical and the laboratory experiments, the physical conditions
of this benchmark are considered.
The length of the channel is $6$ meters, and its sketch can
be seen in Figure \ref{fig:channel-sketch}. The initial condition is water at rest
with $\eta=0$ and a triangular block of sediments, whose geometry is depicted in Figure \ref{fig:channel-sketch}.
In our numerical results, we will have four  
 tide-gauges, $N=4$, where laboratory measures have been taken each $5$ milliseconds,
thus generating four tidal series. 
These buoys are located at the positions $1.87$, $2.87$,
$3.87$ and $4.87$ meters, and they are depicted in Figure \ref{fig:channel-sketch}. 
We take $g=9.81$ $m/s^2$ and $L=5$ layers of fluid in the model.

The calibration tests are run until $T=8$ seconds both for the synthetic test and 
the laboratory experiment.
For the finite volume method we consider $200$ cells in the analytical test and $800$ cells
in the laboratory essay with $CFL=0.5$.

We recall that 
the parameters are three, $\pmb p=(r,\theta, n)$,
where $r$ is the ratio of densities between the fluid and the sediment, $\theta$
the Coulomb angle, and $n$ the friction coefficient. The search domain 
for all the experiments in this section is 
$D=[0.3,0.8]\times [5,45]\times [10^{-5},10^{-3}]$,
which is quite a broad domain.

Concerning the hardware configuration,
all tests have been performed in a 
server with $16$ CPU cores
(two Intel Xeon E5-2620 v4 clocked at 2.10GHz, accounting 32 logical threads) and $16$ GB of RAM.

\begin{figure}[!htb]
\begin{center}
\includegraphics[width=11cm]{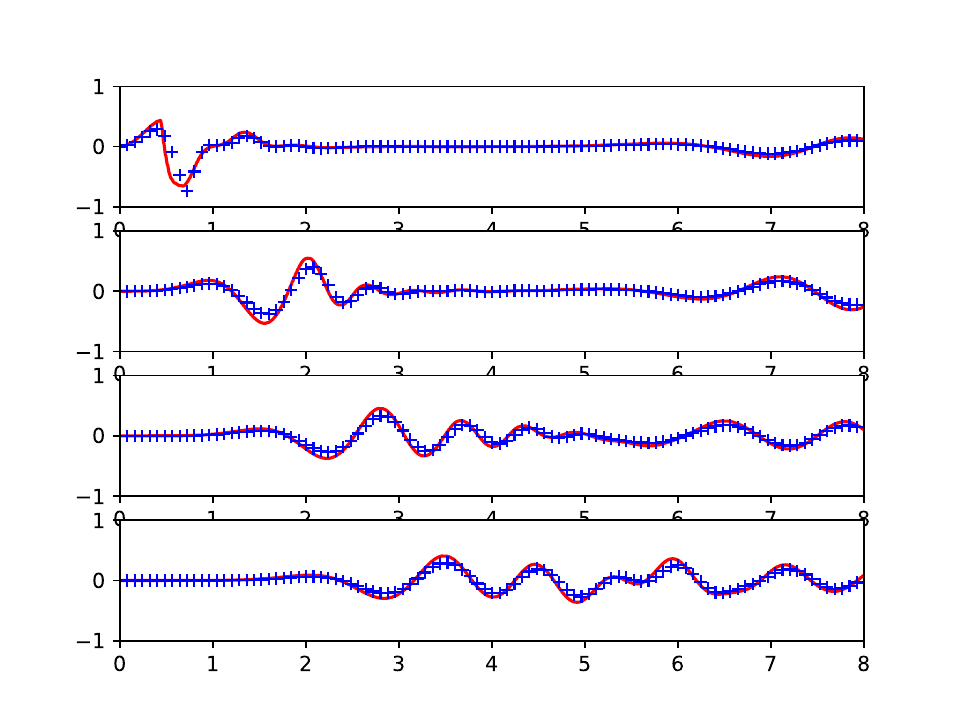}
\end{center}
\caption{\red{Synthetic generated series vs calibrated ones with the multi-start L-BFGS-B. Target series in red, simulated series in blue.}}
\label{fig:test_local1_analytic}
\end{figure}

\begin{table}[!htb]
\begin{center}
\begin{tabular}{c|c|c|c}
 & $r$ & $\theta$ & $n$ \\
\hline
Target values & $0.55$ & $12\degree$ & $0.0002$ \\
\hline
Obtained values & $0.55$ & $12\degree$ & $0.0002$ 
\end{tabular} 
\caption{\label{tb1} Target and obtained values of the parameters.}
\end{center}
\end{table}

\begin{figure}[!htb]
\begin{center}
\includegraphics[width=13cm]{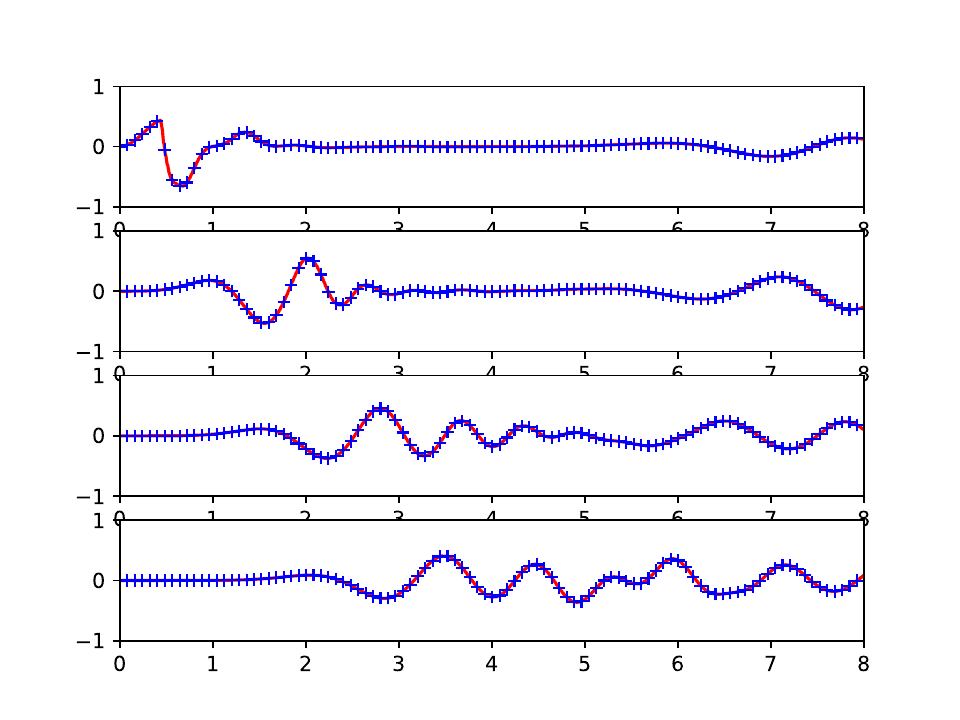}
\end{center}
\caption{Synthetic generated series vs calibrated ones. Target series in red, simulated series in blue.\label{fig:analytical-vs-callibrated-series}}
\end{figure}

\subsection{Synthetic test\label{sec:analytical}}

\red{

The notion of identifiability addresses the question of whether it is at all possible to obtain unique solutions of the inverse problem for unknown parameters of interest in a model from data collected in the spatial and temporal domains \cite{Navon,Cacuci13}. As we have seen, data assimilation problems deal in the end with the search of the global minimum of a cost function. The exploration of the global minimum is a nontrivial task as long as the cost function has a complicated structure, and, on top of that, ensuring that the involved cost function has a unique global minimum is a extremely difficult goal, mainly due to the fact that sophisticated numerical methods are needed to simulate from landslide tsunami models being able to recover realistic physical phenomena, as discussed in Section \ref{sec:model}.

Analyzing parameter identifiability is  precisely one of the aims of this work. We seek to check whether the data assimilation problem for landslide tsunami models is well posed when using only information of the fluid free surface.
In fact, our goal in this first set of numerical experiments is precisely to  empirically discuss the \blueJA{problem} of identifiability and uniqueness of the here proposed parameters calibration strategy. We will observe that the parameters are identifiable using only data from the free surface, which is something unexpected and eye catching from our point of view, because at first sight one could expect that information of the lower layers, the sediment layer or  the speed of the fluid should be required in order to assimilate the data into the model. Nevertheless, in practice the information of the free surface proofs to be enough.

In this work, as in the article \cite{Sumata}, the uniqueness of the minimum of the cost function will be discussed by invoking results from the following additional optimization experiments.

}

\blueJA{
\subsubsection{Synthetic test 1}
}

\red{
First of all, we designed a synthetic experiment, where given the unknown set of parameters, we created the observations numerically, which were then assimilated into the model to retrieve the original set of parameters. }
The values of the parameters were set at 
$r=0.55$,  $\theta=12\degree$ and  $n=0.0002$. The test was run for $8$ seconds.
With these data we computed the simulation and stored the series corresponding with the
free surface at each measure point in an interval of $0.005$ seconds. Then, we supposed that the 
parameters were unknown and tried to recover them using our optimization
algorithms.
\red{
There is no doubt about the uniqueness of the global minimum:
the value of the cost function at this unique global minimum is zero, since the observations are perfect because they arise from the model.} 
This problem has a very similar 
 level of complexity 
from the optimization
point of view to the real one we want to tackle, although it has the
advantage of
being easier to handle, as 
 the exact solution is known.
Moreover, this benchmark allowed us to test and compare the different
algorithms with different number of parallel search paths.

First we show  that if a local optimization algorithm, like  L-BFGS-B is applied, which can be seen as performing only one
path and one temperature step of the hybrid algorithm, no convergence to global minimum is obtained. Thus, after executing a local L-BFGS-B searcher, starting from a random point of the search domain,
the obtained set of parameters is $(r,\theta,n)=(6.826989\times 10^{-01} ,  10.68841753^o, 8.178492\times 10^{-4})$,
the value of the cost function being $5.273542\times 10^{-02}$. The simulation obtained with 
this set of parameters is shown  and compared with the exact solution at Figure \ref{fig:test_local1_analytic}. Therefore, a robust global optimization algorithm should be used to compute the global minimum of this problem.

\begin{table}[!ht]

\centering
\begin{tabular}{|c|c|r|r|r|}
\hline
\multirow{16}{*}{\begin{sideways}Simulated Annealing\end{sideways}} & \#Threads & ${\mathcal T}$ & \#Func Evals & Cost Function \\
\cline{2-5}
 & \multirow{3}{*}{1} & $1$ & $101$ & $3.25\times 10^{-2}$  \\
 \cline{3-5}
 &  & $0.48$ & $ 3.61 \times 10^3$ & $2.31 \times 10^{-2}$  \\
 \cline{3-5}
 &  & $10^{-4}$ & $3.92 \times 10^5$ & $1.11 \times 10^{-3}$  \\
 \cline{2-5}
 & \multirow{3}{*}{2} & $1$ & $201$ & $3.51 \times 10^{-2}$  \\
 \cline{3-5}
 &  & $0.48$ & $7.21 \times 10^3$ & $2.16 \times 10^{-2}$  \\
 \cline{3-5}
 &  & $10^{-4}$ & $7.83 \times 10^5$ & $4.58 \times 10^{-4}$  \\
 \cline{2-5}
 & \multirow{3}{*}{4} & $1$ & $401$ & $2.38 \times 10^{-2}$  \\
 \cline{3-5}
 &  & $0.48$ & $1.44 \times 10^4$ & $1.82 \times 10^{-2}$  \\
 \cline{3-5}
 &  & $10^{-4}$ & $1.57 \times 10^6$ & $9.11 \times 10^{-5}$  \\
 \cline{2-5} 
 & \multirow{3}{*}{8} & $1$ & $801$ & $1.73 \times 10^{-2}$  \\
 \cline{3-5}
 &  & $0.48$ & $2.88 \times 10^4$ & $1.24 \times 10^{-2}$  \\
 \cline{3-5}
 &  & $10^{-4}$ & $3.13 \times 10^6$ & $1.33 \times 10^{-4}$  \\
 \cline{2-5} 
 & \multirow{3}{*}{16} & $1$ & $1601$ & $7.88 \times 10^{-3}$ \\
 \cline{3-5}
 &  & $0.48$ & $5.76 \times 10^4$ & $7.20 \times 10^{-3}$ \\
 \cline{3-5}
 &  & $10^{-4}$ & $6.27 \times 10^6$ & $3.50\times 10^{-5}$  \\
 \cline{2-5} 
\hline
\multirow{16}{*}{\begin{sideways}Basin Hopping\end{sideways}} & \#Threads & ${\mathcal T}$ & \#Func Evals & Cost Function \\
\cline{2-5}
 & \multirow{3}{*}{1} & $1$ & $82$ & $2.28 \times 10^{-2}$  \\
 \cline{3-5}
 &  & $0.48$ & $1.72 \times 10^3$ & $1.47 \times 10^{-2}$  \\
 \cline{3-5}
 &  & $10^{-4}$ & $1.96 \times 10^5$ & $6.33 \times 10^{-4}$  \\
 \cline{2-5}
 & \multirow{3}{*}{2} & $1$ & $103$ & $2.29 \times 10^{-2}$  \\
 \cline{3-5}
 &  & $0.48$ & $3.48 \times 10^3$ & $1.24 \times 10^{-2}$  \\
 \cline{3-5}
 &  & $10^{-4}$ & $3.15 \times 10^5$ & $1.50 \times 10^{-3}$  \\
 \cline{2-5}
 & \multirow{3}{*}{4} & $1$ & $285$ & $9.83 \times 10^{-3}$  \\
 \cline{3-5}
 &  & $0.48$ & $6.99 \times 10^3$ & $2.71 \times 10^{-3}$  \\
 \cline{3-5}
 &  & $10^{-4}$ & $6.49 \times 10^5$ & $1.33 \times 10^{-4}$ \\
 \cline{2-5} 
 & \multirow{3}{*}{8} & $1$ & $369$ & $2.08 \times 10^{-2}$ \\
 \cline{3-5}
 &  & $0.48$ & $1.30 \times 10^4$ & $ 7.04 \times 10^{-4}$ \\
 \cline{3-5}
 &  & $10^{-4}$ & $1.38 \times 10^6$ & $1.04 \times 10^{-4}$ \\
 \cline{2-5} 
 & \multirow{3}{*}{16} & $1$ & $825$ & $1.17 \times 10^{-2}$ \\
 \cline{3-5}
 &  & $0.48$ & $2.51 \times 10^4$ & $3.75 \times 10^{-4}$  \\
 \cline{3-5}
 &  & $10^{-4}$ & $2.67 \times 10^6$ & $9.92\times 10^{-6}$  \\
 \cline{2-5} 
\hline

\end{tabular}
\caption{Parallel SA (SA$_\text{M}$) vs. parallel BH (BH$_\text{M}$). The column labeled as ``Cost Function'' shows the value of the cost function at the best point visited so far by the minimization algorithm.}
\label{tab:SAvsBH}

\end{table}

\begin{table}[!htb]
\begin{center}
\begin{tabular}{c|c|c|c|c|c}
Number of cores & 1 & 2 & 4 & 8 & 16 \\
\hline
Time (seconds) & $872.64$ & $1640.57$ & $3035.07$ & $5493.48$ & $9338.92$\\
\hline
Speedup & 1   &  $1.88$ & $3.47$ & $6.30$ & $10.70$ \\
\end{tabular} 
\caption{Multi-path BH$_{16}$: speedup using multi-CPU implementation.}
\label{table:speedup-multi-pathBH}
\end{center}
\end{table}

Figure \ref{fig:analytical-vs-callibrated-series} shows the results obtained if both hybrid multi-path SA or BH algorithm are applied. Now, the parameters are computed exactly (see Table \ref{tb1}), and a perfect agreement between the signals is observed.

\begin{figure}[!htb]
\begin{center}
\includegraphics[width=11cm]{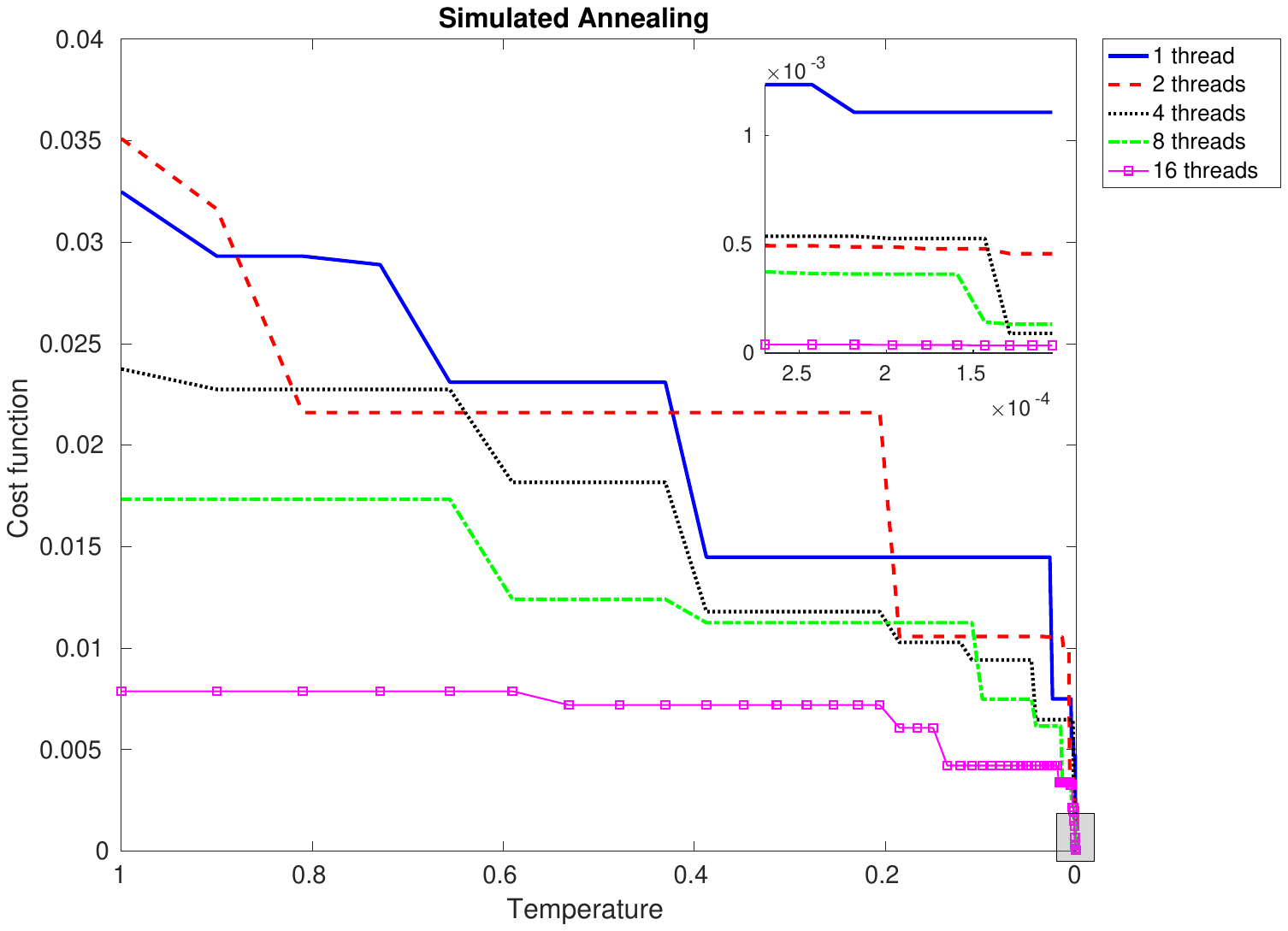}
\end{center}
\caption{Convergence of multi-path SA, with 1,2,4,8 and 16 search paths.}
\label{fig:temperatureSA}
\end{figure}

\begin{figure}[!htb]
\begin{center}
\includegraphics[width=11cm]{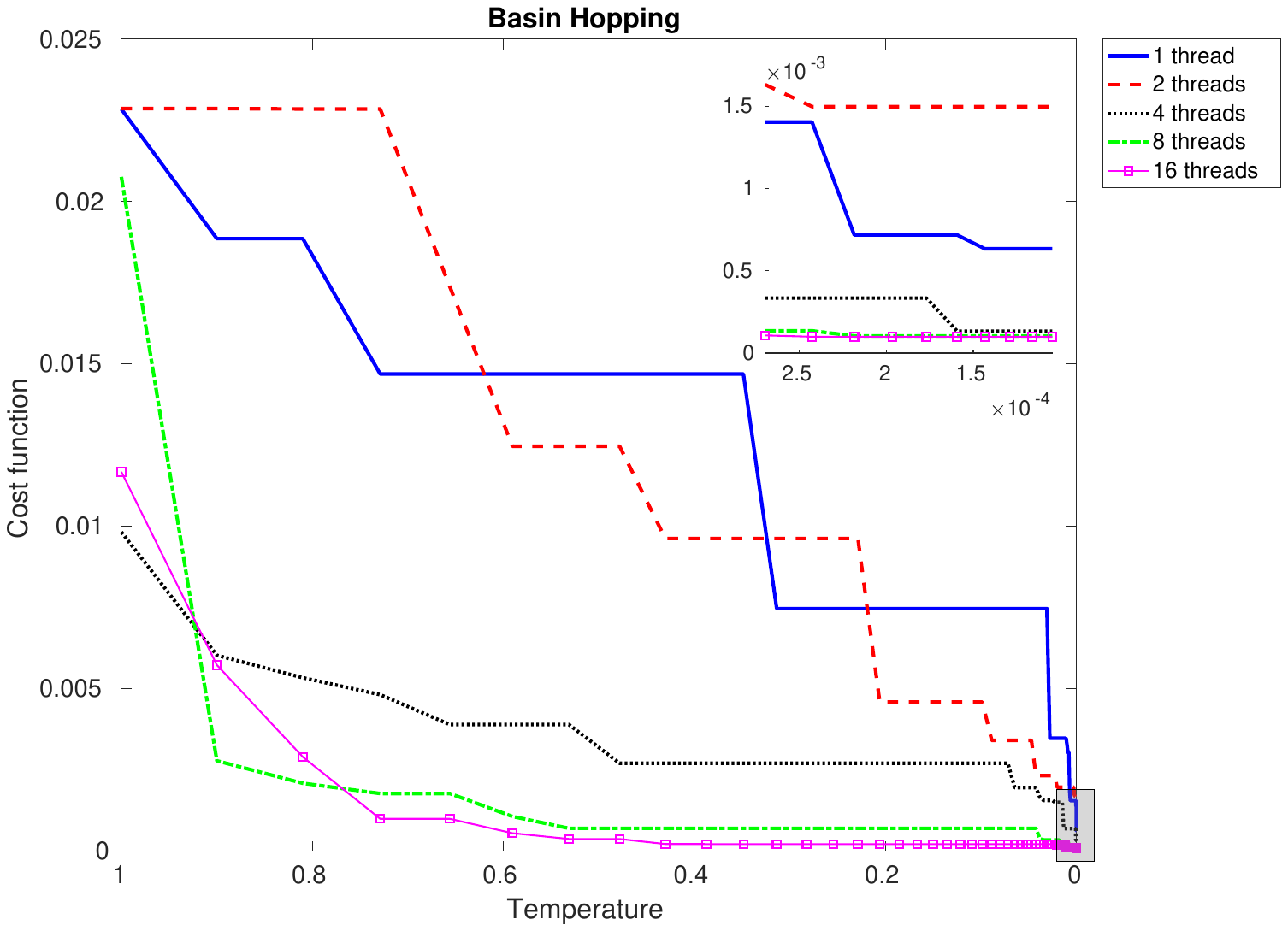}
\end{center}
\caption{Convergence of multi-path BH, with 1,2,4,8 and 16 search paths.}
\label{fig:temperatureBH}
\end{figure}

We can also use this benchmark to assess the convergence and efficiency of the 
two proposed hybrid multi-path global optimization algorithms. In Figures  \ref{fig:temperatureSA} and \ref{fig:temperatureBH},  we show a comparison of the convergence of 
SA$_\text{M}$ and BH$_\text{M}$ algorithms, respectively, using different number of paths ranging from 
$1$ to $16$. At each temperature, the value of the cost function at the best point visited so far by the algorithm is shown. Note that the current state of the minimizer at each stage could be different to the referred best visited point owing to the stochastic nature of the SA and BH algorithms.

In Table \ref{tab:SAvsBH} we show the convergence of the multi-path algorithms when increasing the 
number of paths. The convergence speed is shown in terms of the number of function evaluations performed by the algorithm. This number of evaluations is shown at different levels of
temperatures in the annealing process, ${\mathcal T}=1, 0.48, 10^{-4}$, and for different number of search paths, ranging from $1$ to $16$.
The computing time of each evaluation for this test is $~2.8$ seconds in our hardware configuration.
We want to remind that when doing more than one search, the searches are distributed among the number of CPU cores, and that for BH$_\text{M}$ this number
of evaluations include the three extra evaluations performed for computing the gradients.
In Table \ref{table:speedup-multi-pathBH} we show the parallel computational
efficiency, in termns of the speedup, when using multiple cores
for performing  $16$ search paths,
with a number of threads ranging from $1$ to $16$.

\begin{table}
\begin{center}
\begin{tabular}{c|c|c|c|c}
 & \multicolumn{3}{c|}{Parameters} & \\
 \hline
Gauges & $r$ & $\theta$ & $n$ & Cost func. \\
\hline
$G1$-$G2$-$G3$-$G4$ & $0.55$ & $12^o$ & $2\times 10^{-4}$ & 
$9.923\times 10^{-6}$\\
\hline
$G3$-$G4$ & $5.493439\times 10^{-1}$ & $ 11.8404944^o$ & $2.045592\times 10^{-4}$ &  
$6.234\times 10^{-4}$ \\  
\hline
$G4$ & $5.529343\times 10^{-1}$ & $11.2507678^o $ & $2.132503\times 10^{-4}$ &
 $1.648\times 10^{-3}$\\  
\end{tabular} 
\caption{Obtained values of the parameters and value of cost function.}
\label{tab:obtained-parameters-analytic-test}
\end{center}
\end{table}

\begin{center}
\begin{figure}[htbp]
\red{
\begin{center}
\includegraphics[width=0.8\textwidth]{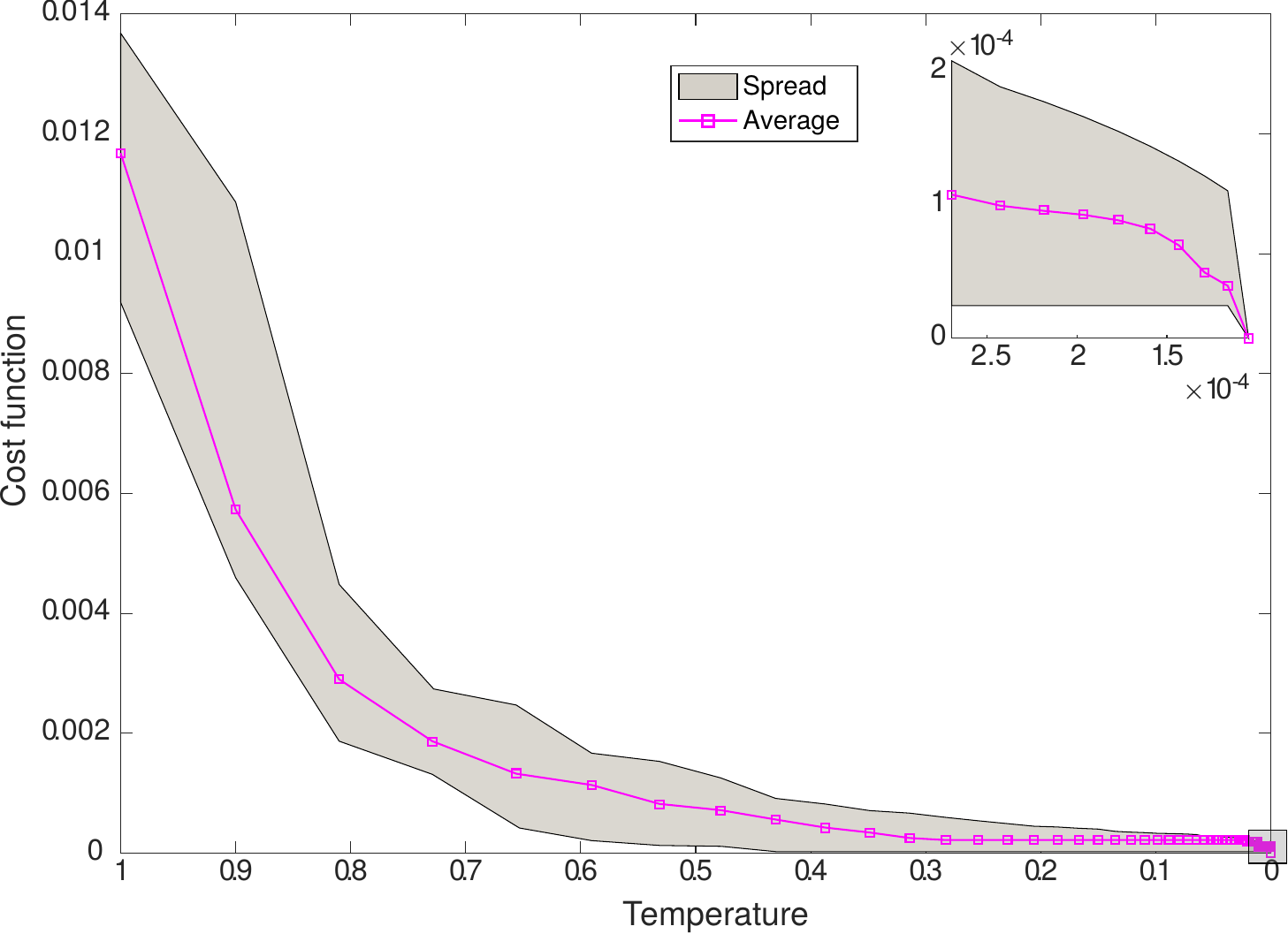} 
\caption{Evolution of the cost function for $20$ optimization experiments. The gray shade denotes the spread (the range of maximum-minimum cost) and the squared line the average of the $20$ experiments.} 
\label{fig:spreads}
\end{center}
}
\end{figure}
\end{center}

\begin{table}[h]
\red{
\begin{center}
\begin{tabular}{|r|r|r||r|r|}
\hline
$r$ & $\theta$ & $n$ & Cost func.   \\
\hline
\hline
 $0.35$ & $ 12$ & $0.0002$ & $2.19\times 10^{-6}$ \\ %
\hline
$0.35 $ & $12$ & $0.0004$ & $ 9.14 \times 10^{-6}$   \\ %
\hline
 $0.35$ & $25$ & $0.0002$ & $ 2.99\times 10^{-6}$  \\ %
\hline
$0.35$ & $25$ & $0.0004$ & $ 1.53\times 10^{-6}$  \\ %
\hline
$0.35$ & $37$ & $0.0002 $ & $ 3.80 \times 10^{-6}$  \\ %
\hline
$0.35$ & $37$ & $0.0004$ & $6.40 \times 10^{-6}$  \\ %
\hline
$0.55$ & $12$ & $0.0002$ & $ 2.89\times 10^{-6}$  \\ %
\hline
$0.55$ & $12$ & $0.0004 $ & $8.65 \times 10^{-6}$  \\ %
\hline
$ 0.55$ & $25$ & $0.0002  $ & $6.83 \times 10^{-6}$  \\ %
\hline
$0.55$ & $25$ & $0.0004 $ & $4.75 \times 10^{-6}$  \\ %
\hline
$ 0.55$ & $37$ & $0.0002 $ & $2.76 \times 10^{-6}$  \\ %
\hline
$ 0.55$ & $37$ & $0.0004$ & $8.36 \times 10^{-6}$  \\ %
\hline
$ 0.75$ & $12$ & $0.0002$ & $2.49 \times 10^{-6}$  \\ %
\hline
$ 0.75$ & $12$ & $0.0004$ & $8.71 \times 10^{-6}$  \\ %
\hline
$ 0.75$ & $25$ & $0.0002$ & $4.87 \times 10^{-6}$  \\ %
\hline
$ 0.75$ & $25$ & $0.0004 $ & $5.96 \times 10^{-6}$  \\ %
\hline
$ 0.75$ & $37$ & $0.0002$ & $4.72 \times 10^{-6}$  \\ %
\hline
$ 0.75$ & $37$ & $0.0004$ & $2.29 \times 10^{-6}$  \\ %
\hline
\end{tabular}
\end{center}
\caption{Values of the cost function for several data assimilations.
\label{table:identifiability}}
}
\end{table}

Next, we check the convergence of the algorithm to the global optimum when a  lower number of measure points is used. We made the experiment of considering
the time series of the free surface only at tide-gauge G4, or only at tide-gauges G3-G4.
In Table \ref{tab:obtained-parameters-analytic-test}
we show the value of the cost function together with the obtained set of parameters using only data from G4,
and 
the same information when calibrating against tide-gauges G3-G4. 
As expected, the value of the cost function is better when taking the four tide-gauges.

\begin{figure}[h!tbp]
\red{
\centering
 \subfigure[$r=0.3$, $\theta=12$, $n=0.0002$.]{\includegraphics[scale=0.41]{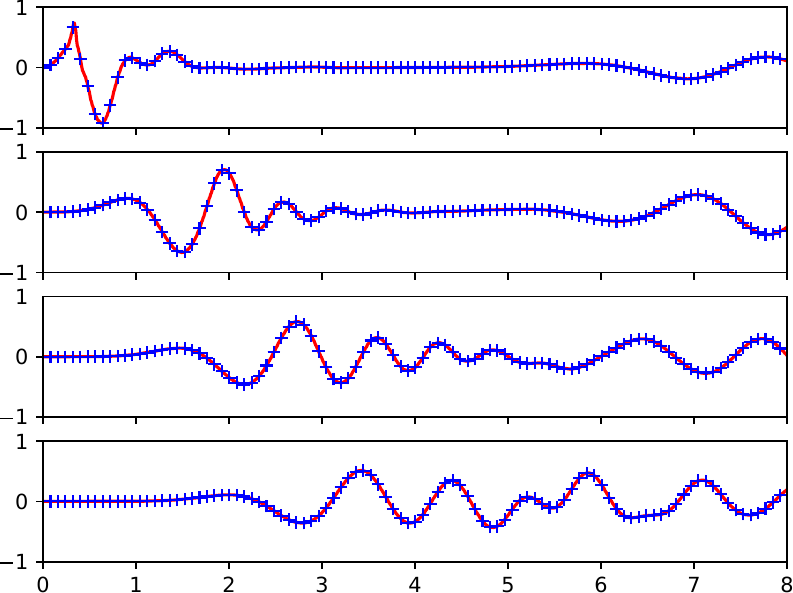}}
 \subfigure[$r=0.5$, $\theta=22$, $n=0.0002$.]{\includegraphics[scale=0.41]{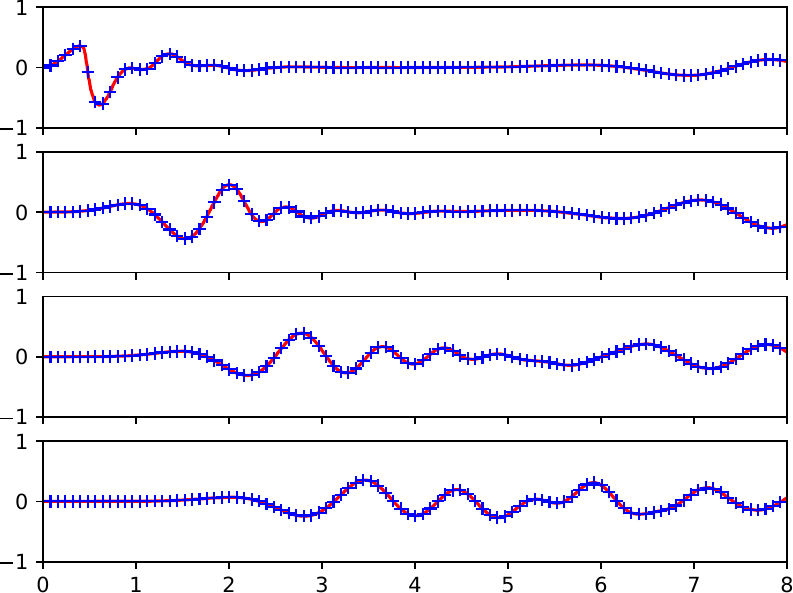}}
 
 \subfigure[$r=0.6$, $\theta=31$, $n=0.0004$.]{\includegraphics[scale=0.41]{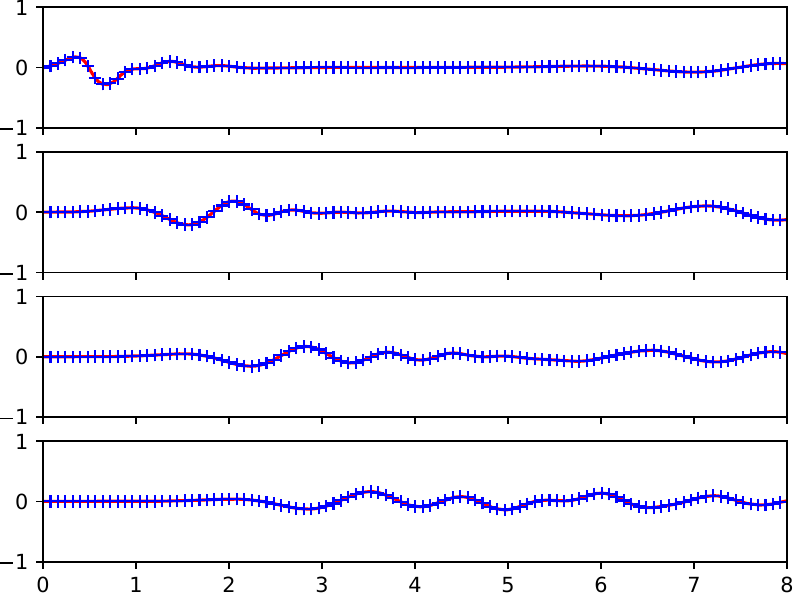}} 
 \subfigure[$r=0.8$, $\theta=39$, $n=0.0004$.]{\includegraphics[scale=0.41]{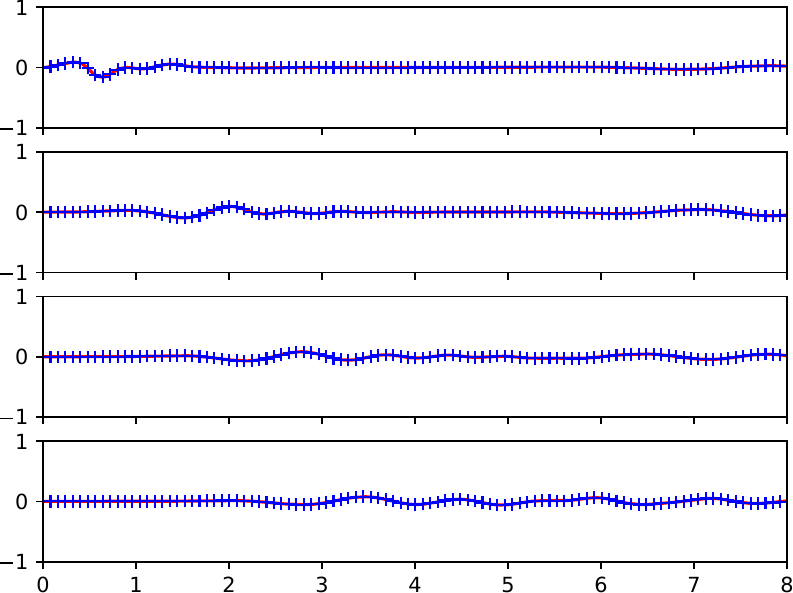}}   
\caption{Synthetic generated series vs calibrated ones. In red, target series that are a priori generated with a set of known parameters. In blue, simulated series obtained with the assimilated parameters achieved with the global optimizer.} 
\label{fig:identifiability}
}
\end{figure}

\blueJA{
\subsubsection{Synthetic test 2}
}
\red{
Secondly, we ran a pool of $20$ independent optimization experiments with our set up, each optimization starting from different initial parameter values, randomly chosen in the search domain, and each test used a different seed for the creation of the random numbers consumed by the algorithm in order to explore  the search domain, i.e. each experiment performed a seek of the minimum from a different starting point along a different search path. 
Figure 
\ref{fig:spreads}  shows the evolution of the cost function and its spread for the $20$ optimization experiments. The spread is defined by the range of the maximum  value of the cost function and its minimum in the set of the $20$ optimizations at each temperature step. The average of the $20$ realizations was also computed. More precisely at each temperature, the worse, the best, and the average of the best points visited by each one of the twenty minimizers up to the current temperature, are shown.
All experiments show an asymptotic reduction of the values of the cost function toward the same zero value, and none of the optimizations ends up in a local minimum.  Therefore, this study clearly shows that this stochastic approach (hybrid local-global optimization) is suitable to find the global minimum of a structurally complicated cost function.

\bigskip
\blueJA{
\subsubsection{Synthetic test 3}
}

Finally, we sampled the search domain with $18$ sets of parameters, for all of them, once more we generated the corresponding synthetic tests and performed a \blueJA{successful} data assimilation, these experiments being summarized on Table \ref{table:identifiability}. We note that with this pool of data we swimmingly calibrated the model to all types of waves varying from those with very high amplitudes to the flat ones, see Figure \ref{fig:identifiability}. Notice that this smoothing effect was obtained by increasing more and more the ratio of densities $r$ and the Coulomb angle $\theta$. Therefore, the issue of identifiability is accomplished for the very different types of possible waves.
}

\bigskip

\subsection{Application to a laboratory test with real 
data\label{sec:experimental}}

\begin{figure}[!htb]
\begin{center}
\includegraphics[width=9cm]{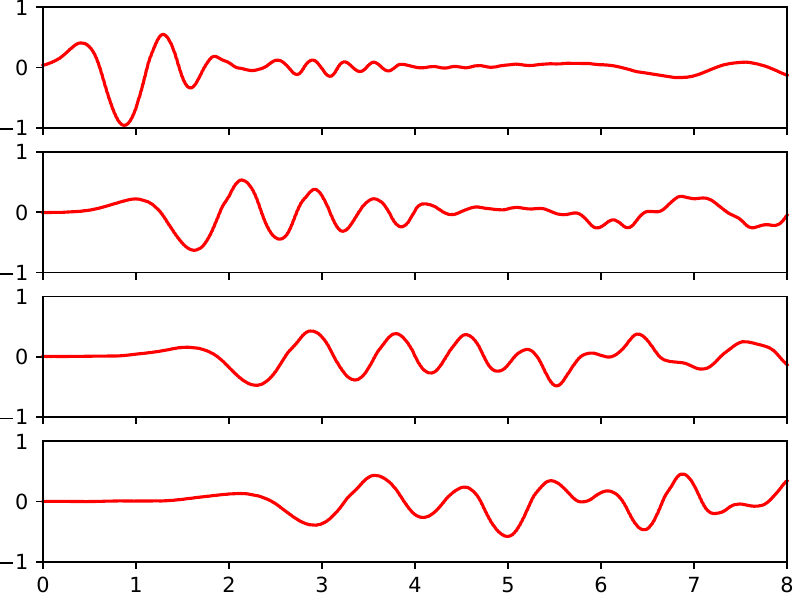}
\end{center}
\caption{Series mearured in laboratory experiments.}
\label{fig:laboratory-time-series-initial}
\end{figure}

In this experiment we performed the data assimilation for a real situation
where laboratory series of the free surface for four measure points were given. The experiment
was performed at \'Ecole Centrale de Marseille (IRPHE), France, \cite{Grilli2017}. The positions of the measure buoys were once more
$1.87$, $2.87$, $3.87$ and $4.87$ meters. The time series for these points are shown in 
Figure \ref{fig:laboratory-time-series-initial}. These time series, together with the
description of the experiment and 
some videos, are available in the web page \cite{web-experiments}.

\begin{figure}[!htb]
\begin{center}
\includegraphics[width=11cm]{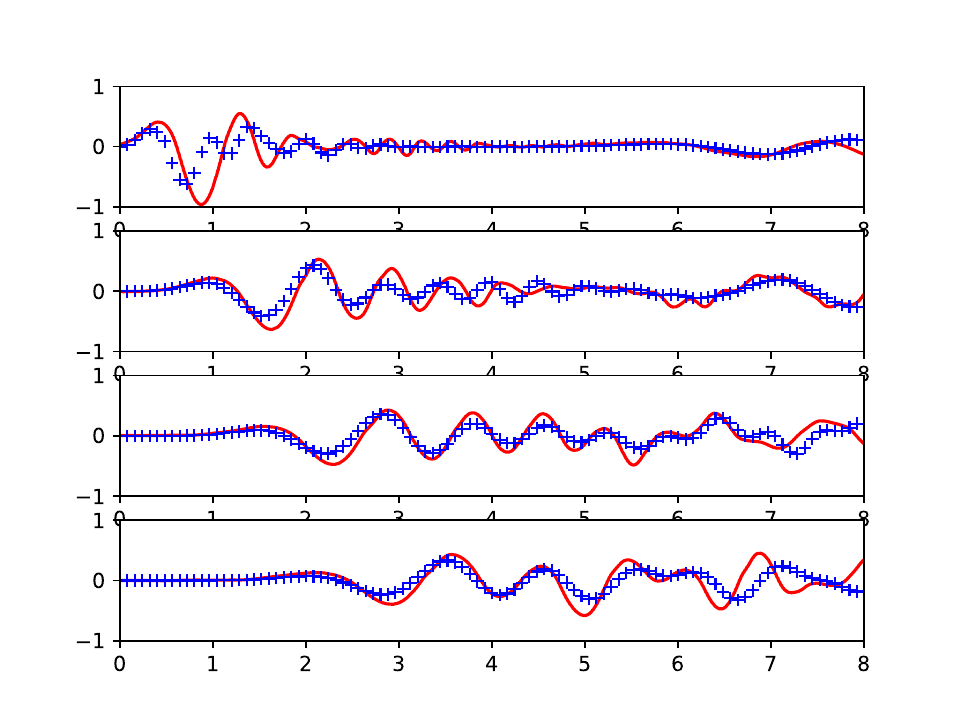}
\end{center}
\caption{Multi start solution with $32$ L-BFGS-B local searches: computed signals in blue and laboratory data in red.}
\label{fig:test_local_analytic}
\end{figure}

One more time, first we show that the results obtained with a multi-start algorithm are worse than those obtained with a hybrid multi-path algorithm.
For example, if we apply a multi-start  L-BFGS-B to this problem, the obtained solution does not match adequately the laboratory data (see Figure \ref{fig:test_local_analytic}). 
This experiment corresponds with
launching only one temperature step of BH$_\text{M}$ with $32$ paths, and the set of 
obtained parameters is 
$(r,\theta,n)=(4.632742\times 10^{-01} , 9.48064720^o , 8.176018\times 10^{-1})$, for which the value of the cost function is $1.321301\times 10^{-01}$. 

\begin{center}
\begin{figure}[!htb]
\begin{center}
\includegraphics[width=12cm]{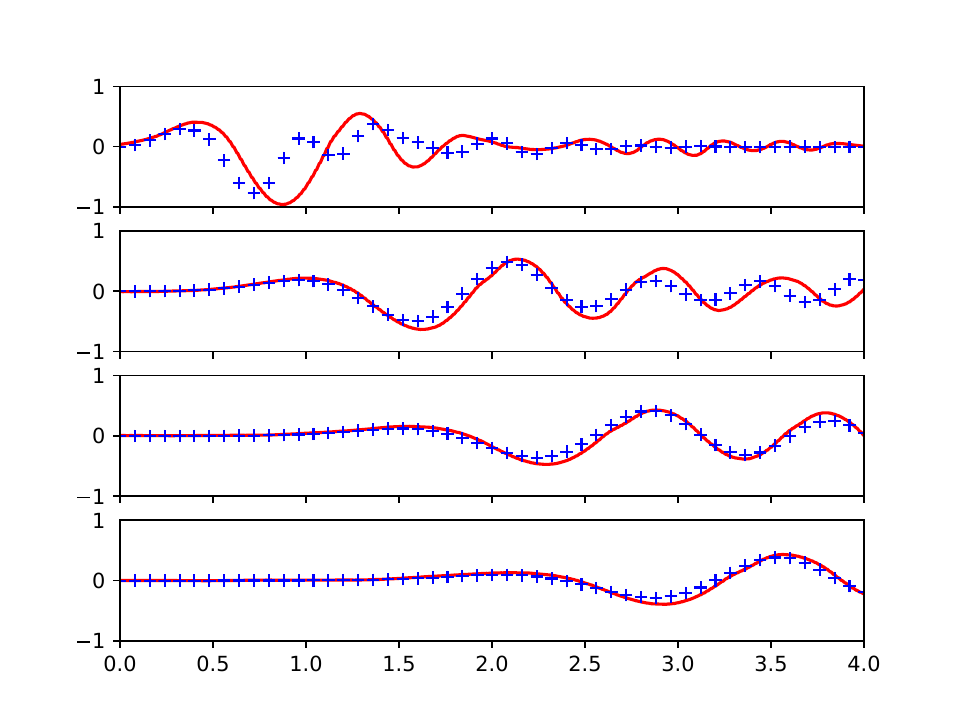}
\caption{Laboratory series vs calibrated ones. Lab series in red, simulated series in blue. From top to bottom, free surface at tide-gauges $G1$, $G2$, $G3$ and $G4$.
\label{fig:series-lab-vs-callibrated-4}}
\end{center}
\end{figure}
\end{center}

\begin{center}
\begin{figure}[!htb]
\begin{center}
\includegraphics[width=12cm]{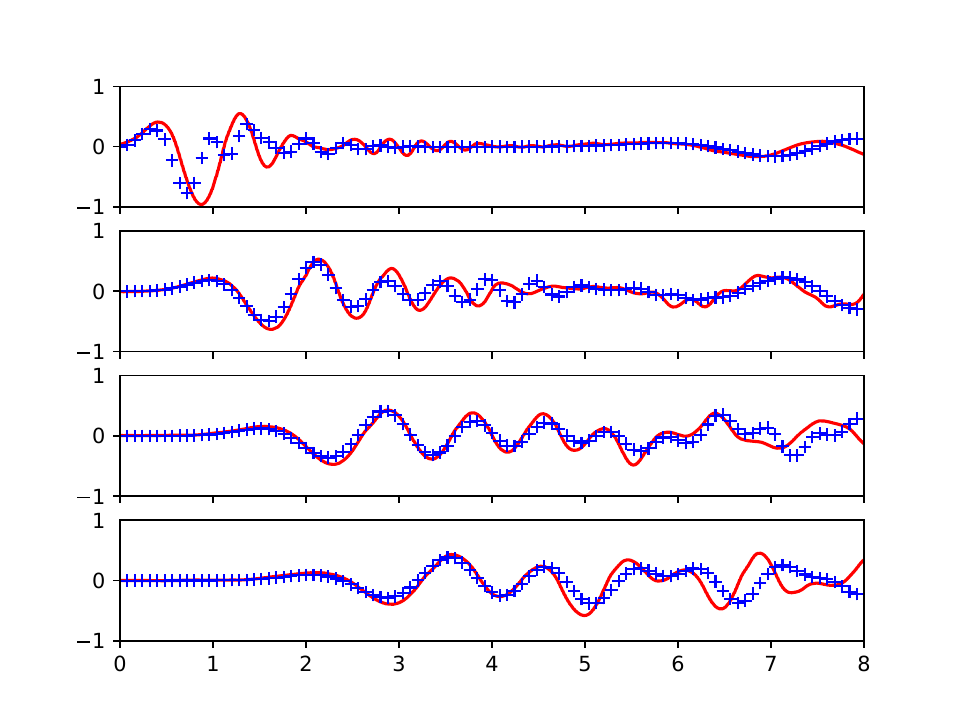}
\caption{Laboratory series vs calibrated ones. Lab series in red, simulated series in blue.  From top to bottom, free surface at tide-gauges $G1$, $G2$, $G3$ and $G4$.
\label{fig:series-lab-vs-callibrated-8}} 
\end{center}
\end{figure}
\end{center}

After global calibration with BH$_\text{M}$, the results can be seen in Figures \ref{fig:series-lab-vs-callibrated-4} 
and  \ref{fig:series-lab-vs-callibrated-8}.  The obtained values for the parameters are shown in Table \ref{tab:obtained-parameters-lab-test} and the value of the cost function is $1.224355\times 10^{-1}$.

In those figures we can see that with the calibrated
set of parameters a good agreement in the signals amplitudes and pulses is obtained, between laboratory and simulated data. 
The approximation is even better up to the 4th second (see Figure \ref{fig:series-lab-vs-callibrated-4}). The matching is quite good
at initial seconds, and it becomes worse as time evolves. Also we see a better agreement 
for the farthest tide-gauges, $G3$ and $G4$, and it becomes worse the closer we are to the
the initial position of the landslide, close to tide-gauge $G1$.
The amplitudes of the signal are very well captured by the model.
The period (pulses, maximums and minima of the signal) is  well captured for the 
last three tide-gauges until the fourth second, and there is a little gap
from that time on. The first tide-gauge is difficult to be captured by the model.
Further investigation should be done. In fact, at this early stage, compaction and dilatancy effects are quite important, and they  are not taken into account in the here considered landslide model. Therefore, a more accurate model for the landslide motion is needed to better simulate this early stages of the landslide motion.

Newly in this laboratory experiment we repeated the practice of using a lower number of measure points. We made the test of considering
the free surface series only at tide-gauge $G4$, or only at tide-gauges $G3$-$G4$,
the results can be seen in Table \ref{tab:obtained-parameters-lab-test}.
The obtained error considering the four series until time $T=8$ seconds, using the parameters calibrated with 
only the last tide-gauge $G4$, is $1.31589\times 10^{-1}$. 
Besides, the obtained error considering the four series, using the parameters calibrated with 
only the last two tide-gauges $G3$-$G4$ is $1.24279\times 10^{-1}$. 
The result is  not too poor when considering only the measures of the last tide-gauge, nevertheless it is off course much better when considering
$G3$-$G4$.
Using the last two tide-gauges, the free surface series are quite close to the best obtained result using the four tide-gauges, and
also interesting is the fact that the set of parameters gets closer to the ones obtained with the four tide-gauges.

\begin{table}
\begin{center}
\begin{tabular}{c|c|c|c|c}
 & \multicolumn{3}{c|}{Parameters} & \\
 \hline
Gauges & $r$ & $\theta$ & $n$ & Cost func. \\
\hline
$G1$-$G2$-$G3$-$G4$ & $0.6501164$ & $6.03510265^o$ & $4.3690\times 10^{-4}$ & 
$1.224355 \times10^{-1}$\\
\hline
$G3$-$G4$ & $7.080885\times 10^{-1}$ & $5.38770216^o$ & $3.144702\times 10^{-4}$ &  
$1.24279 \times10^{-1}$ \\  
\hline
$G4$ & $7.633579\times 10^{-1}$ & $5.16240342^o $ & $2.397312\times 10^{-4}$ &
 $1.31589 \times10^{-1}$\\  
\end{tabular} 
\caption{Obtained values of the parameters and value of cost function.}
\label{tab:obtained-parameters-lab-test}
\end{center}
\end{table}

\section{Conclusions}

We have shown that hybrid multi-path global optimization 
algorithms can be suitable for solving the data assimilation problem
for models of submarine avalanches. 

\blueJA{
Besides, we have assessed  the identifiability of the model,
if only data of the free surface is available,
i.e.  we have checked that
the data assimilation problem is well posed when calibrating 
only against  measures of the fluid free surface.}

We have discussed that using a local optimizer 
or a multi-start techique produces poor results, and that the consideration of global optimization algorithms is more suitable for this kind of problems. We have also exhibited that the problem can be solved using gradient numerical
optimization algorithms in the local part. 

This calibration procedure/technique results also interesting because it allows to measure 
the quality of the model: the quality of two different models can be 
quantitative (not only qualitative) compared attending to the result of the 
calibration. 
It provides us with a machinery 
for comparing the good properties of different models. The one with the lowest 
minimum, can be quantitative said to better approximate the real physical problem.

The laboratory experiment is quite challenging. The obtained results look promising, although a perfect match between laboratory data and the calibrated model has not been achieved due to limitations of the underlying model. In any case, we have shown that the  multi-path BH algorithm could be used to calibrate this kind of problems. Moreover, this opens the door to the use of this global optimization machinery for real problems, and in particular, for helping in developing better  models for landslide tsunamis and assessing their precision and adjustment to the laboratory data.

\section{Acknowledgements}
The authors want to acknowledge the designers of the experiment (\cite{MA201540}), for making the data publicly available. \red{The authors also wish to thank the anonymous reviewers for their through review of the article and their constructive advises.}

This research has been financially supported by Spanish Government Ministerio de Econom\'ia y Competitividad through the research projects MTM2016-76497-R and MTM2015-70490-C2-1-R.

\bibliography{bibfile}

\end{document}